\documentclass[a4paper,USenglish,cleveref,autoref,numberwithinsect]{lipics-v2021} 

\usepackage{csquotes}
\usepackage{microtype}

\usepackage{subcaption}
\newtheorem{mythm}{Theorem}
\newtheorem{myq}{Open Problem}
\newtheorem{myhope}{Conjecture}
\newtheorem{myobs}[mythm]{Observation}
\newtheorem{mylem}[mythm]{Lemma}
\newtheorem{mycor}[mythm]{Corollary}

\crefname{figure}{Figure}{Figures}
\crefname{mythm}{Theorem}{Theorems}
\crefname{myhope}{Conjecture}{Conjectures}

\hideLIPIcs
\nolinenumbers

\title{Separable Drawings: Extendability and Crossing-Free Hamiltonian Cycles} 

\author{Oswin Aichholzer}{Institute of Algorithms and Theory, Graz University of Technology, Austria}{oswin.aichholzer@tugraz.at}{https://orcid.org/0000-0002-2364-0583}{Partially supported by the Austrian Science Fund (FWF) grant W1230.}

\author{Joachim Orthaber}{Institute of Algorithms and Theory, Graz University of Technology, Austria\\ Institute of Mathematics, Technische Universit{\"a}t Berlin, Germany}{orthaber@math.tu-berlin.de}{https://orcid.org/0000-0002-9982-0070}{Supported by the Austrian Science Fund (FWF) grant W1230.}

\author{Birgit Vogtenhuber}{Institute of Algorithms and Theory, Graz University of Technology, Austria}{birgit.vogtenhuber@tugraz.at}{https://orcid.org/0000-0002-7166-4467}{Partially supported by Austrian Science Fund within the collaborative DACH project \emph{Arrangements and Drawings} as FWF project \mbox{I 3340-N35}.}

\authorrunning{O. Aichholzer, J. Orthaber, and B. Vogtenhuber}

\ccsdesc[500]{Mathematics of computing~Combinatorics}
\ccsdesc[500]{Mathematics of computing~Graph theory}
\ccsdesc[500]{Theory of computation~Computational geometry}

\keywords{Simple drawings, Pseudospherical drawings, Generalized convex drawings, Plane Hamiltonicity, Extendability of drawings, Recognition of drawing classes}

\relatedversion{A preliminary version of this paper appeared at GD 2024~\cite{aov-2024-sdecfhc-gd}.}

\acknowledgements{We thank Alexandra Weinberger for fruitful discussions during the early stages of our research on this topic.
We further thank all the anonymous referees for their helpful comments on the paper.}

\begin{document}

\maketitle

\begin{abstract}
Generalizing pseudospherical drawings, we introduce a new class of simple drawings, which we call \emph{separable drawings}.
In a separable drawing, every edge can be closed to a simple curve that intersects each other edge at most once.
For different edges, the non-edge parts of these curves may interact arbitrarily though.
Most notably, we show that
(1)~every separable drawing of any graph on $n$ vertices in the plane can be extended to a simple drawing of the complete graph $K_n$,
(2)~every separable drawing of $K_n$ contains a crossing-free Hamiltonian cycle and is plane Hamiltonian connected (that is, it contains a crossing-free Hamiltonian path between each pair of vertices), and
(3)~every generalized convex drawing and every 2-page book drawing is separable.
Further, the class of separable drawings is a proper superclass of the union of generalized convex and 2-page book drawings.
Hence, our results on plane Hamiltonicity extend recent work on generalized convex drawings by Bergold et al.~(DCG 2025).
\end{abstract}

\section{Introduction}\label{sec:intro}

A \emph{simple drawing} of a graph $G$ is a representation of $G$ in the plane (or on the sphere) such that the vertices of $G$ are mapped to distinct points and the edges of $G$ are mapped to Jordan arcs connecting their respective end-vertices.
Furthermore, every pair of edges is allowed to have at most one point in common, which is either a common end-vertex or a proper crossing.
Simple drawings of graphs are widely studied combinatorial objects that have received considerable attention in different areas of graph drawing.
They are important for the crossing number~\cite{bls-2019-cihc,gj-1983-cnnpc,prtt-2006-iclfmcsg} because every crossing-minimizing drawing of a graph is simple~\cite{schaefer-2018-cng}, and, apart from the topics in this paper, they have also been studied, for example, in the context of empty triangles~\cite{ahprsv-2015-etgdcg,gtvw-2023-etgtdkn,h-1998-etdcg,r-2015-etctg} and thrackles~\cite{fp-2019-tiub,ps-2011-ccmt,x-2021-nubct}.

Very special and intensively studied simple drawings are \emph{straight-line drawings}, where the edges are the straight-line segments between their end-vertices.
A fundamental difference between these two drawing classes is that straight-line drawings are defined geometrically (restricting the shape of the edges) while simple drawings are defined combinatorially (restricting crossing properties).
To bridge this gap, several classes in between have been considered.
One possibility is to relax the geometric restriction on the edges and consider, for example, \emph{$x$-monotone} \cite{ahpsv-2015-dmgdcg,aov-2024-tcfhcsdcg,ks-2025-esmd-iwoca} or \emph{c-monotone drawings} \cite{agtvw-2024-twfpssdcg,f-2014-endestgcd,r-2017-mdetg}, where the edges are monotone with respect to one direction or one point in the plane, respectively.%
\footnote{For more details on and relations between these and several more classes of simple drawings (some of which are mentioned later) we refer the interested reader to an earlier paper of ours~\cite{aov-2024-tcfhcsdcg}.}
Alternatively, we can directly switch to a combinatorially defined class such as \emph{pseudolinear drawings} \cite{abr-2021-edgap,s-2021-tdgtpdcg}, which slightly generalize straight-line drawings and have been studied for a long time~\cite{l-1926-tpegp}.
For these drawings there exists an arrangement of pseudolines (bi-infinite curves that pairwise cross exactly once) such that every edge lies on one pseudoline.

Generalizing the latter, a drawing class that was recently introduced by Arroyo, Richter, and Sunohara~\cite{ars-2021-edcgap} and is of special interest for this work is the class of \emph{pseudospherical drawings}.
These are simple drawings $\mathcal{D}$ in which every edge $e$ is contained in a simple closed curve $\gamma_e$ such that
\begin{enumerate}
\item\label{ps:endvtcs} the only two vertices of $\mathcal{D}$ on $\gamma_e$ are the end-vertices of $e$,
\item\label{ps:pc} for any two edges $e \neq f$ the curves $\gamma_e$ and $\gamma_f$ intersect in exactly two crossing points, and
\item\label{ps:onecrossing} $\gamma_e$ intersects every edge $f\neq e$ of $\mathcal{D}$ at most once, either in a crossing or in an end-vertex.
\end{enumerate}
The idea behind these three properties is that they combinatorially generalize \emph{spherical drawings}, where the vertices are points on the sphere and the edges are geodesics.

Further generalizing pseudospherical drawings, we introduce a new class of simple drawings which we call \emph{separable drawings}.
These are all simple drawings that fulfill Properties~\ref{ps:endvtcs} and~\ref{ps:onecrossing} of pseudospherical drawings, but not necessarily Property~\ref{ps:pc}.
Separable drawings can also be seen as \enquote{locally pseudospherical} because locally for every edge, they behave like pseudospherical drawings, but the non-edge parts of curves $\gamma_e$ and $\gamma_f$ for different edges $e$ and $f$ in $\mathcal{D}$ may interact arbitrarily.
This additional freedom gives the advantage that for recognizing separable drawings, it is sufficient to find a curve for each edge of the drawing independently.
That is, we do not have to consider the set of potential curves for all edges simultaneously, which can be relevant from a computational point of view.
Moreover, we show that it is in fact additional freedom in the sense that the class of separable drawings is strictly larger than the one of pseudospherical drawings.

Our starting point for coming up with separable drawings was the following question brought to our attention by Bruce Richter (see also \cite{amrs-2022-cdcgtmg,ars-2021-edcgap}):
Is every crossing-minimizing drawing of the complete graph $K_n$ pseudospherical?
Simultaneously, we were studying two classic graph drawing questions for simple drawings, namely, the extendability to simple drawings of complete graphs and the existence of crossing-free Hamiltonian cycles in drawings of complete graphs.
In this work, we answer both these questions for the class of separable drawings, we elucidate the relation of separable drawings to further classes of simple drawings, and we resolve the question of how hard it is to recognize a separable drawing.
Before we go into more detail about our contribution, we give some background information on the considered problems.

\subparagraph*{Edge Extension of Simple Drawings.}

It is easy to see that every straight-line drawing in the plane on $n$ vertices in general position can be extended to a straight-line drawing of the complete graph~$K_n$.
As a consequence of Levi's Enlargement Lemma~\cite{l-1926-tpegp}, an analogous statement is true for pseudolinear drawings.
Recently, Kyn{\v{c}}l and Soukup~\cite{ks-2025-esmd-iwoca} showed that also every $x$-monotone drawing admits an extension to an $x$-monotone drawing of the complete graph.
However, for simple drawings the situation is very different.
Kyn\v{c}l showed that extendability to complete graphs is not always possible~\cite{k-2013-iestg}.
Even more, there exist simple drawings of graphs that have only a linear number of edges but cannot be extended by any of the missing edges without violating simplicity~\cite{hirs-2018-ss2stgfe}.
The decision problem of whether a given drawing can be extended by some set of given edges is NP-complete~\cite{adp-2019-esd}, even when the set consists of a single edge~\cite{akpsvw-2023-ioesdh}.
Remarkably, this NP-hardness still holds when restricting to \emph{pseudocircular drawings}, which are all simple drawings that fulfill Properties~\ref{ps:endvtcs} and~\ref{ps:pc} of the definition of pseudospherical drawings.
On the positive side, the edge extension problem is fixed-parameter tractable (FPT) when parameterized by the number of edges to insert and an upper bound on newly created crossings~\cite{ghkpv-2021-coesd}.
The complexity of deciding whether a given simple drawing can be extended to a simple drawing of the complete graph is still an open problem.
We remark that the above-mentioned hardness results for extension via a fixed edge do not directly imply the hardness of extendability to the complete graph.

\subparagraph*{Crossing-Free Hamiltonian Cycles and Paths.}

A subdrawing of a simple drawing is \emph{crossing-free} (also known as \emph{plane}) if no two edges in that subdrawing cross each other.
It is well known that every straight-line drawing of $K_n$ contains a crossing-free Hamiltonian cycle, while this property does not hold anymore for straight-line drawings of general graphs.
Indeed, it already breaks for $K_n$ minus one edge:
Place the vertices in convex position and remove one edge from its convex hull.
In 1988, however, Rafla~\cite{r-1988-gdcg} conjectured a generalization to simple drawings for the case of complete graphs.

\begin{myhope}[\cite{r-1988-gdcg}]\label{conj:cycles}
Every simple drawing of $K_n$ with $n \geq 3$ vertices contains a crossing-free Hamiltonian cycle.
\end{myhope}

If \Cref{conj:cycles} is true, then every simple drawing of $K_n$ also contains at least $n$ crossing-free Hamiltonian paths and, for even $n$, at least $2$ crossing-free perfect matchings.
Pach, Solymosi, and T{\'o}th~\cite{pst-2003-ucctg} popularized the study of crossing-free subdrawings.
For simple drawings, a lot of effort went into the search for crossing-free matchings \cite{fs-2009-dtbgrrtr,f-2014-endestgcd,pt-2010-detg,r-2017-mdetg,s-2013-dectg}, with the current best lower bound for their size being $\Omega(\sqrt{n})$~\cite{agtvw-2024-twfpssdcg}.
With regard to special drawing classes, the existence of a crossing-free Hamiltonian cycle in \emph{$2$-page book drawings}\footnote{Note that 2-page book drawings are drawings of \emph{linear layouts} with 2 partition classes~\cite{dw-2004-llg}.} (where all vertices lie on a straight line and the edges are drawn as half-circles) or even in $x$-monotone drawings can be established with basic tools.
Further, \Cref{conj:cycles} was proven to hold for generalized twisted drawings on an odd number of vertices~\cite{agtvw-2024-twfpssdcg}.
In a previous work, we also confirmed it for cylindrical drawings as well as strongly c-monotone drawings~\cite{aov-2024-tcfhcsdcg}.
In that work, we further stated the following conjecture, which we showed to be a strengthening of \Cref{conj:cycles} for all simple drawings of complete graphs, but not necessarily for a restricted class of simple drawings.
Moreover, we proved that for cylindrical and for strongly c-monotone drawings, fulfilling the property of \Cref{conj:paths} actually implies the property of \Cref{conj:cycles} and we confirmed both conjectures restricted to these classes.

\begin{myhope}[\cite{aov-2024-tcfhcsdcg}]\label{conj:paths}
Every simple drawing $\mathcal{D}$ of $K_n$ with $n \geq 2$ vertices contains, for each pair of vertices $v \neq w$ in $\mathcal{D}$, a crossing-free Hamiltonian path with end-vertices $v$ and $w$.
\end{myhope}

Very recently, both conjectures have also been verified restricted to the large class of g-convex\footnote{%
G-convex drawings are just called convex drawings in~\cite{amrs-2022-cdcgtmg,bfmos-2025-phccd}. However, we prefer the term g-convex (short for generalized convex) to avoid confusion, since the term convex drawing classically refers to a straight-line drawing with vertices in convex position.}
drawings~\cite{bfmos-2025-phccd}; see below for its definition.
In their paper, the authors further coined the term \emph{plane Hamiltonian connected} for drawings fulfilling the property of \Cref{conj:paths}.
Moreover, they showed a relevantly stronger result for the class of g-convex drawings of~$K_n$, namely, that every spanning star in such a drawing is \emph{compatible} with some crossing-free Hamiltionian cycle.
Compatibility of two crossing-free subdrawings means that also their union is crossing-free, a concept that has been studied for example with regard to matchings~\cite{abdghhkmrssuw-2009-cgm,abls-2015-bcm,ist-2013-dcgm,prss-2020-aggm-gd} and spanning trees~\cite{akmeopmvw-2023-cstsdkn-gd,ghht-2014-cst}.
The authors of the before-mentioned work~\cite{bfmos-2025-phccd} additionally certified with some examples that the compatibility of spanning stars with crossing-free Hamiltionian cycles does not hold anymore for general simple drawings of~$K_n$.
However, they conjectured the following, which combines plane Hamiltonicity with results on maximal plane subdrawings~\cite{gpt-2021-psctd,r-2015-etctg}.

\begin{myhope}[\cite{bfmos-2025-phccd}]\label{conj:2n-3}
Every simple drawing of $K_n$ with $n \geq 3$ vertices contains a crossing-free subdrawing with at least $2n-3$ edges that in turn contains a Hamiltonian cycle.
\end{myhope}

A simple drawing $\mathcal{D}$ of $K_n$ is called \emph{g-convex} if every \emph{triangle} in $\mathcal{D}$ has a \emph{convex side}.
A~\emph{triangle} in $\mathcal{D}$ is the simple closed curve formed by an induced subdrawing on three vertices in~$\mathcal{D}$.
Every triangle splits the plane (or sphere) into two connected components, their closures are the \emph{sides} of the triangle.
A side $S$ of a triangle is called \emph{convex} if the subdrawing of $\mathcal{D}$ that is induced by all vertices in $S$ is completely contained in $S$ (that is, no edge between two such vertices crosses the triangle).
Note that both sides of a triangle in a g-convex drawing can be convex.

G-convex drawings have been introduced by Arroyo, McQuillan, Richter, and Salazar~\cite{amrs-2022-cdcgtmg} as the largest class of a hierarchy of drawing classes that bridges the gap between straight-line drawings and simple drawings of~$K_n$ by successively relaxing combinatorial properties.
Hence the results on plane Hamiltonicity in g-convex drawings~\cite{bfmos-2025-phccd} are quite strong.

\subparagraph*{Recognition and Realizability.}

Two simple drawings of a graph are \emph{strongly isomorphic} if they can be transformed into each other via a homeomorphism of the sphere.
Arroyo, Bensmail, and Richter~\cite{abr-2021-edgap} showed that the question of recognizing whether a simple drawing of an arbitrary graph is strongly isomorphic to a pseudolinear drawing can be solved in polynomial time (their result is for a given drawing in the plane; for isomorphism on the sphere it is sufficient to consider every face as the unbounded one).

Two simple drawings of a graph are \emph{weakly isomorphic} if they have the same crossing edge pairs.
Clearly any two strongly isomorphic drawings are also weakly isomorphic while the converse is in general not true.
Both concepts group the infinite number of drawings of a fixed graph into a finite number of equivalence classes.
A third such concept are rotation systems, which are a classic combinatorial abstraction of simple drawings.%
\footnote{Rotation systems are also an abstraction of the more general concept of star-simple drawings~\cite{aeppv-2017-ssdcg-egc,fhkkp-2022-mncssdknnel}.}
The \emph{rotation} of a vertex in a simple drawing is the (clockwise) cyclic order of its incident edges, which is classically given by an accordingly sorted list of its adjacent vertices.
The \emph{rotation system} of a simple drawing is the collection of the rotations of all of its vertices.
Gioan~\cite{g-2022-cgdtm} and Kyn{\v{c}}l~\cite{k-2011-srcatgp} independently showed that two simple drawings of $K_n$ are weakly isomorphic if and only if they have the same rotation system.
Hence, recognizing whether a simple drawing of $K_n$ is weakly isomorphic to a simple drawing of a certain class can be accomplished by showing that its rotation system has a realization as such a drawing.

An \emph{abstract rotation system} of $K_n$ gives, for each vertex, a (potentially arbitrary) cyclic order of its incident edges.
As shown by Kyn{\v{c}}l~\cite{k-2020-srcatgs} in combination with computational results from~\cite{aafhpprsv-2015-agdscg}, an abstract rotation system is realizable as a simple drawing if and only if all its subrotation systems on five vertices are.
Consequently, this can be checked in $\mathcal{O}(n^{5})$ time.

Aichholzer, Hackl, Pilz, Salazar, and Vogtenhuber~\cite{ahpsv-2015-dmgdcg} showed that deciding whether a rotation system of $K_n$ can be realized as an $x$-monotone or a $2$-page book drawing can also be done in $\mathcal{O}(n^{5})$ time.
Arroyo, McQuillan, Richter, and Salazar~\cite{amrs-2022-cdcgtmg} showed that an according statement holds for realizability as a g-convex drawing, by showing that g-convex drawings are exactly those drawings in which every $5$-tuple of vertices admits a straight-line realization.
On the other hand, deciding whether a rotation system is realizable as a straight-line drawing is known to be $\exists \mathbb{R}$-complete, since it can be reduced to the realizability of abstract order types and the stretchability of pseudoline arrangements; see for example the survey by Schaefer, Cardinal, and Miltzow~\cite{scm-2024-etrccc}.

\subparagraph*{Our Contribution.}

We start in \Cref{sec:basics} by introducing some more notation and showing first properties of separable drawings, also motivating why we chose the name \enquote{separable}.
We further observe that every $2$-page book drawing is separable (\Cref{obs:2-page}).

In \Cref{sec:extendability} we consider the extension problem.
We prove that for every graph $G$ on $n$ vertices, every separable drawing of $G$ can be completed to a simple drawing of~$K_n$ (\Cref{thm:extend-separable}).
This is remarkable since the problem is NP-complete for the closely related class of pseudocircular drawings (recall that the class of pseudospherical drawings is the intersection of the classes of separable and pseudocircular drawings).
We further discuss that extension to simple drawings is the best we can hope for by presenting an example of a separable drawing that cannot be extended to any separable drawing of~$K_n$ (\Cref{fig:non-sep-extend}).
Moreover, we show that analogous statements hold for crossing-minimizing drawings of~$G$ (\Cref{thm:extend-cross-min,fig:crossing-minimal}).
By this we establish a potential connection between separable and crossing-minimizing drawings.

In \Cref{sec:hamiltonian}, we turn our attention to the plane Hamiltonicity problem.
We show that all separable drawings of $K_n$ fulfill the properties of both \Cref{conj:cycles} (\Cref{thm:cycle-separable}) and \Cref{conj:paths} (\Cref{thm:paths-separable}).
In the process we also confirm the property of \Cref{conj:2n-3} for them.
Moreover, we prove that separable drawings are a proper superclass of g-convex drawings (\Cref{thm:gconvex}).
Thus our results on plane Hamiltonicity constitute a strengthening of the according results on g-convex drawings~\cite{bfmos-2025-phccd}.
As a distinction between the two classes, we observe that the stronger compatibility of spanning stars with crossing-free Hamiltonian cycles in g-convex drawings does not transfer to separable drawings (\Cref{obs:separable-star-no-compatible-hc}).

In \Cref{sec:flips} we introduce flips in rotation systems and use them to show that, for simple drawings of~$K_n$, being separable is a property of the rotation system (\Cref{thm:separator-flip-equivalence}).
Finally, we consider the question of recognizing separable drawings in \Cref{sec:recognition}.
This is of particular interest in case it should turn out that crossing-minimizing drawings are separable.
We show that the recognition problem is solvable in polynomial time for simple drawings of~$K_n$ (\Cref{thm:poly}) but NP-complete for simple drawings of general graphs (\Cref{thm:npc}).

We conclude with several open problems in \Cref{sec:conclusion}.

\section{Basic Properties}\label{sec:basics}

Before we get to first properties of separable drawings, we introduce some more notation to facilitate argumentation.
We call an edge $e$ of a simple drawing $\mathcal{D}$ a \emph{separator edge} if there exists a simple closed curve $\gamma_{e}$ containing $e$ such that the only vertices of $\mathcal{D}$ on $\gamma_{e}$ are the end-vertices of $e$ and such that, for each edge $f \neq e$ of $\mathcal{D}$, $\gamma_{e}$ has at most one point in common with~$f$.
We call the curve~$\gamma_{e}$ a \emph{witness} for $e$.
With this definition, a simple drawing~$\mathcal{D}$ is separable if and only if every edge of $\mathcal{D}$ is a separator edge.

Note that a simple closed curve $\gamma$ partitions the plane into two connected components.
We call the closures of these components the \emph{sides} of~$\gamma$.
To ease reasoning, we sometimes refer to the bounded side of $\gamma$ in the plane as the \emph{inside} and the other side as the \emph{outside}.

The following lemma motivates why we call separable drawings \enquote{separable}.

\begin{mylem}\label{lem:basic-separation}
Let $\gamma_{e}$ be a witness of a separator edge $e$ in a simple drawing~$\mathcal{D}$.
Then every edge $f$ of~$\mathcal{D}$ that connects two vertices on the same side of $\gamma_{e}$ is fully contained in that side.
\end{mylem}

\begin{proof}
The statement is clear for $e$ itself.
Further, by the definition of a separator edge, each edge $f \neq e$ of $\mathcal{D}$ has at most one point in common with~$\gamma_{e}$.
Every edge $f$ incident to $e$ already has an end-vertex in common with $\gamma_e$ and, therefore, is contained in one side of~$\gamma_e$.
Finally, every edge $f$ with both end-vertices on the same side of $\gamma_e$ and not incident to $e$ crosses $\gamma_e$ an even number of times.
Since $f$ crosses $\gamma_e$ at most once, it does not cross $\gamma_e$ at all, which implies that $f$ is contained in one side of~$\gamma_e$.
\end{proof}

\Cref{lem:basic-separation} tells us that, for every edge $e$ in a separable drawing $\mathcal{D}$, each witness $\gamma_e$ of~$e$ separates $\mathcal{D}$ into two induced subdrawings, one lying inside of $\gamma_e$ and one outside, that together cover all vertices of~$\mathcal{D}$ and only share the common edge~$e$.
For simple drawings of~$K_n$, as we show next, the existence of two such induced subdrawings is actually an equivalent characterization of separability.
This implies that, for complete graphs, we do not need to check edges between the two sides of $\gamma_e$ for multiple intersections with~$\gamma_e$.

\begin{mylem}\label{lem:separator-easier-kn}
Let $\mathcal{D}$ be a simple drawing of $K_{n}$ and let $e = \{ v, w \}$ be an edge of $\mathcal{D}$. Then $e$ is a separator edge if and only if $e$ can be closed to a simple curve $\gamma'_e$ such that every edge $f$ of $\mathcal{D}$ that connects two vertices on the same side of $\gamma'_{e}$ is fully contained in that side.
\end{mylem}

\begin{proof}
The \enquote{only if} part is true by \Cref{lem:basic-separation}.
For the other direction, let $\mathcal{D}_1$ be the subdrawing of $\mathcal{D}$ induced by all vertices of the inside of $\gamma'_e$ and similarly $\mathcal{D}_2$ for all vertices of the outside of $\gamma'_e$.
Since $\gamma'_e$ separates $\mathcal{D}_1$ and $\mathcal{D}_2$, any edge in $\mathcal{D}_1$ or $\mathcal{D}_2$ has at most one point in common with~$\gamma'_e$.
It remains to consider edges $f = \{ v_1, v_2 \}$ with $v_1 \in \mathcal{D}_1$ and $v_2 \in \mathcal{D}_2$.
If $f$ crosses $e$, which it can cross at most once, then $f$ lies in the inside of the crossing $K_4$ on the vertices $\{ v, w, v_1, v_2 \}$.
In other words, $f$ is separated from $\gamma'_e \setminus e$ by the $4$-cycle $(v,v_1,w,v_2)$.
Hence it cannot cross $\gamma'_e$ a second time.
The other case is that $f$ crosses $\gamma'_e \setminus e$.
Let $\mathcal{B}$ be the boundary of the unbounded cell of~$\mathcal{D}_1$.
We show that $f$ crosses $\mathcal{B}$ exactly once.

Assume for a contradiction that  $f$ crosses $\mathcal{B}$ more than once. Let $x_{1}$ and $x_{2}$ be two consecutive of these crossings along $f$ such that the part $f'$ between $x_{1}$ and $x_{2}$ lies inside~$\mathcal{B}$.
Then $f'$ crosses $\mathcal{D}_1$ and separates the inside of $\mathcal{B}$ into two connected components $F_1$ and~$F_2$.
Let $F_1$ be the component that contains $e$.
If no vertex of $\mathcal{D}_1$ lies in $F_2$, then every edge in~$\mathcal{D}_1$ that is crossed by $f'$ would have to be crossed at least twice, a contradiction to $\mathcal{D}$ being simple.
Hence, there is some vertex $z$ of $\mathcal{D}_1$ in~$F_2$.  
If $v_{1}$ lies in $F_1$, then $\{ v_{1}, z \}$ crosses its incident edge~$f$.
If $v_{1}$ lies in $F_2$, then $\{ v_{1}, v \}$ and $\{ v_{1}, w \}$ cross their incident edge $f$.
Since we have a contradiction in both cases, $f$~crosses $\mathcal{B}$ exactly once.

We reroute $\gamma'_e \setminus e$ arbitrarily close to $\mathcal{B}$ along the outside of $\mathcal{B}$.
This does not change any crossings with $\mathcal{D}_1$ or $\mathcal{D}_2$ and, by the arguments above, every edge $f$ between the two subdrawings $\mathcal{D}_1$ or $\mathcal{D}_2$ is crossed exactly once by the adapted curve $\gamma'_e$.
Consequently, $e$ is a separator edge.
\end{proof}

Note that the subdrawings $\mathcal{D}_1$ and $\mathcal{D}_2$ in the proof are interchangeable. That is, we could also reroute $\gamma_e$ close to the boundary of~$\mathcal{D}_2$.

A special case of a separator edge is an uncrossed edge~$e$.
Indeed, we can close $e$ to a simple curve $\gamma_{e}$ in a small neighborhood of $e$ itself.
Then $\gamma_e$ has one point in common with every edge incident to $e$ and no point in common with any other edge.
With respect to the separation into two subdrawings, this means that one of them only consists of the edge~$e$.

\begin{myobs}\label{obs:uncrossed}
Every uncrossed edge is a separator edge.
\end{myobs}

Before we come to the main results of this paper, let us mention 2-page book drawings~$\mathcal{D}$.
There the vertices lie on a common line and all edges $e$ are drawn as half-circles.
Hence, by closing $e$ to a circle, we get a simple closed curve that has at most one point in common with any edge $f \neq e$ of~$\mathcal{D}$.

\begin{myobs}\label{obs:2-page}
Every 2-page book drawing is separable.
\end{myobs}

\section{Extendability}\label{sec:extendability}

\begin{figure}[htb]
\centering
\subcaptionbox{\label{fig:reroute-before}}[.32\textwidth]{\includegraphics[page=1]{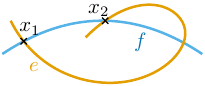}}
\subcaptionbox{\label{fig:reroute-between}}[.32\textwidth]{\includegraphics[page=2]{Figures/reroute.pdf}}
\subcaptionbox{\label{fig:reroute-after}}[.32\textwidth]{\includegraphics[page=3]{Figures/reroute.pdf}}
\caption{\subref{fig:reroute-before}~The edges $e$ and $f$ have more than one point in common, with $x_1$ and $x_2$ being consecutive common points on $f$. \subref{fig:reroute-between}~Every witness $\gamma_g$ that crosses $f_1$ also has to cross $e_1$. \subref{fig:reroute-after}~The result of rerouting $e$ along $f$ between $x_1$ and $x_2$.}
\label{fig:reroute}
\end{figure}

In the following we prove that every separable drawing $\mathcal{D}$ of a graph $G$ on $n$ vertices can be completed to a simple drawing of~$K_n$.
As a first step we show how to add \emph{one} edge to~$\mathcal{D}$.
To do so, we impose a minimality condition regarding the witnesses of all edges in~$\mathcal{D}$.
In particular, we call a collection $\mathcal{D}^{\circ}$ of witnesses, one for every edge in $\mathcal{D}$, a \emph{witness set} for~$\mathcal{D}$.
Further, for an edge $\{ u, v \}$ not in $G$, we call a continuous curve that connects the drawn end-vertices $\mathcal{D}(u)$ and $\mathcal{D}(v)$ in $\mathcal{D}$ a \emph{realization} of $\{ u, v \}$ in $\mathcal{D}$.
Aside from that, we use standard rerouting and exchanging arguments between edges with more than one common point (see \Cref{fig:reroute-before,fig:reroute-after} and \Cref{fig:exchange-before,fig:exchange-after} for illustrations), which remove at least one crossing between the involved edges.

\begin{mylem}\label{lem:extend-separable-by-one}
Let $\mathcal{D}$ be a separable drawing of a non-complete graph $G$ and let $\mathcal{D}^{\circ}$ be a witness set for~$\mathcal{D}$.
For a fixed edge $\{ u, v \}$ not in $G$, let $e$ be a realization of $\{ u, v \}$ in $\mathcal{D}$ that, over all possible realizations, minimizes the number of crossings in $\mathcal{D}^{\circ} + e$.
Then the drawing $\mathcal{D}' = \mathcal{D} + e$ is simple.
\end{mylem}

\begin{proof}
Let $e$ be as described and assume, to the contrary, that $\mathcal{D}'$ is not simple.
The minimality condition implies that $e$ is self-avoiding.
Hence, the non-simplicity assumption implies that $e$ has more than one point in common with some edge $f$ of~$\mathcal{D}$; see \Cref{fig:reroute} for an example illustration.
Let $x_{1}$ and $x_{2}$ be two of those common points that are consecutive along~$f$.
Then the parts $e_{1}$ and $f_{1}$ of $e$ and $f$, respectively, between $x_{1}$ and $x_{2}$ each, together form a simple closed curve.
Since every witness $\gamma_{g}$ in $\mathcal{D}^{\circ}$ for an edge $g$ in $\mathcal{D}$ has at most one point in common with $f$ it follows that, if $\gamma_{g}$ crosses $f_{1}$, then $\gamma_{g}$ also crosses~$e_{1}$ at least once.
Therefore, rerouting $e$ along $f$ between $x_{1}$ and $x_{2}$ reduces the number of crossings of $e$ with $\mathcal{D}^{\circ}$ by at least one and, since $x_{1}$ and $x_{2}$ are consecutive along~$f$, does not introduce self-crossings of~$e$; a contradiction to the minimality condition on~$e$.
\end{proof}

\begin{figure}[htb]
\centering
\subcaptionbox{\label{fig:non-sep-extend-1}}[.32\textwidth]{\includegraphics[page=1]{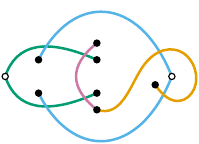}}
\subcaptionbox{\label{fig:non-sep-extend-2}}[.32\textwidth]{\includegraphics[page=2]{Figures/extension.pdf}}
\subcaptionbox{\label{fig:non-sep-extend-3}}[.32\textwidth]{\includegraphics[page=3]{Figures/extension.pdf}}
\caption{\subref{fig:non-sep-extend-1}~A separable drawing of a non-complete graph that cannot be extended to any \emph{separable} drawing of a complete graph. The witnesses of all edges are shown in~\subref{fig:non-sep-extend-2} and \subref{fig:non-sep-extend-3}~shows the only two ways of adding the edge between the two vertices marked with circles.}
\label{fig:non-sep-extend}
\end{figure}

A natural way to get to a simple drawing of $K_n$ would be to iterate the argument of \Cref{lem:extend-separable-by-one}.
However, we would need the drawing in each step to be separable, which might not be the case.
In particular, \Cref{fig:non-sep-extend-1} shows an example of a separable drawing $\mathcal{D}$ on $9$ vertices that cannot be completed to a \emph{separable} drawing of $K_n$.
\Cref{fig:non-sep-extend-2} shows a witness set for $\mathcal{D}$, and \Cref{fig:non-sep-extend-3} indicates that, with respect to crossings, there are only two different ways to add the edge $e$ between the leftmost and rightmost vertex in~$\mathcal{D}$.
Hence the witness of $e$ would have to be the union of these two options.
However, both cross the rightmost edge in~$\mathcal{D}$ (orange), which is not allowed for a witness.
By imposing a second minimality condition, however, we can still extend to a \emph{simple} drawing of~$K_n$.

\begin{figure}[htb]
\centering
\subcaptionbox{\label{fig:exchange-before}}[.32\textwidth]{\includegraphics[page=1]{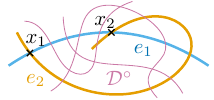}}
\subcaptionbox{\label{fig:exchange-between}}[.32\textwidth]{\includegraphics[page=2]{Figures/exchange.pdf}}
\subcaptionbox{\label{fig:exchange-after}}[.32\textwidth]{\includegraphics[page=3]{Figures/exchange.pdf}}
\caption{\subref{fig:exchange-before}~The edges $e_1$ and $e_2$ have more than one point in common, with $x_1$ and $x_2$ being consecutive common points on $e_1$. \subref{fig:exchange-between}~The parts $e'_1$ and $e'_2$ must have the same number of common points ($4$ each in this example) with the witness set $\mathcal{D}^{\circ}$. \subref{fig:exchange-after}~The result of exchanging $e'_1$ and $e'_2$.}
\label{fig:exchange}
\end{figure}

\begin{mythm}\label{thm:extend-separable}
Let $\mathcal{D}$ be a separable drawing of a non-complete graph on $n$ vertices. Then $\mathcal{D}$ can be extended to a simple drawing of~$K_{n}$.
\end{mythm}

\begin{proof}
Let $\mathcal{D}^{\circ}$ be a witness set for $\mathcal{D}$.
We extend $\mathcal{D}$ to a drawing $\mathcal{D}'$ of $K_{n}$ such that (1)~each added edge $e$ creates a minimum number of additional crossings when being added to $\mathcal{D}^{\circ}$ and such that under this condition (2)~$\mathcal{D}'$ has the least total number of crossings.
Then, by \Cref{lem:extend-separable-by-one}, $\mathcal{D} + e$ is simple for each of the added edges (in particular, all edges are self-avoiding).

Hence, an obstruction to simplicity can only occur between two added edges $e_{1}$ and $e_{2}$ in $\mathcal{D}'$; see \Cref{fig:exchange} for an example illustration.
Let $x_{1}$ and $x_{2}$ be two consecutive common points on $e_1$, and let $e_{1}'$ and $e_{2}'$ be the parts of $e_{1}$ and $e_{2}$, respectively, between $x_{1}$ and~$x_{2}$.
By the first minimality condition, $e_{1}'$ and $e_{2}'$ must have the same number of crossings with~$\mathcal{D}^{\circ}$, otherwise we could reroute either $e_{1}$ or $e_{2}$ along (potentially a part of, to avoid self-crossings) $e_{2}'$ or $e_{1}'$, respectively, to get fewer crossings.
However, then exchanging $e_{1}'$ and $e_{2}'$ (and potentially removing resulting self-crossings on~$e_1$) produces a drawing $\mathcal{D}''$ fulfilling the first minimality condition but with fewer crossings than~$\mathcal{D}'$; a~contradiction to the second minimality condition on~$\mathcal{D}'$.
\end{proof}

Inspired by the question whether every crossing-minimizing drawing of $K_n$ is pseudosphercial, we also investigate and positively answer the extendability of crossing-minimizing drawings of non-complete graphs. 
Interestingly, the proof works rather similarly to that for separable drawings, we only need to replace the arguments regarding the witness set with arguments using that the initial drawing is crossing-minimizing.

Similar to the case of separable drawings, we begin by adding one edge.

\begin{mylem}\label{lem:extend-cross-min-by-one}
Let $\mathcal{D}$ be a crossing-minimizing drawing of a non-complete graph $G$.
For a fixed edge $\{ u, v \}$ not in $G$, let $e$ be a realization of $\{ u, v \}$ in $\mathcal{D}$ that creates a minimum number of additional crossings in $D$.
Then the drawing $\mathcal{D}' = \mathcal{D} + e$ is simple.
\end{mylem}

\begin{proof}
Note that $e$ is clearly self-avoiding and $\mathcal{D}$ is a simple drawing.
Assume now to the contrary that $\mathcal{D}' = \mathcal{D} + e$ is not simple.
Then there exists an edge $f$ in $\mathcal{D}$ such that $e$ and $f$ have more than one point in common.
Let $x_{1}$ and $x_{2}$ be two of those common points that are consecutive along~$f$ and let $\mathcal{D}_{e}^{f}$ be the drawing obtained from $\mathcal{D}'$ by exchanging $e$~and~$f$ between $x_{1}$ and~$x_{2}$.
Then $\mathcal{D}_{e}^{f}$ has one or two crossings less than~$\mathcal{D}'$.
Let, in particular, $c_{1}^{e}$~and $c_{2}^{e}$ be the number of crossings involving $e$, and $c_{1}^{-}$ and $c_{2}^{-}$ the number of crossings not involving $e$ in $\mathcal{D}'$ and $\mathcal{D}_{e}^{f}$, respectively.

If $c_{2}^{-} < c_{1}^{-}$, then removing $e$ from $\mathcal{D}_{e}^{f}$ results in a drawing $\mathcal{D}''$ of $G$ with fewer crossings than $\mathcal{D}$; a contradiction to $\mathcal{D}$ being crossing-minimizing.
Otherwise ($c_{2}^{e} < c_{1}^{e}$), let $\mathcal{D}^{e}$ be the drawing after rerouting $e$ along $f$ between $x_{1}$~and~$x_{2}$~in~$\mathcal{D}'$.
Then, since $e$ does not cross~$f$ between $x_{1}$ and~$x_{2}$, $\mathcal{D}^{e}$~has at most $c_{2}^{e}$ (respectively $+ 1$, if both $x_{1}$ and $x_{2}$ are crossings and the parts of $e$ before $x_{1}$ and after $x_{2}$ lie on different sides of $f$) crossings involving~$e$.
In the $+ 1$ case $\mathcal{D}_{e}^{f}$ has two crossings less than~$\mathcal{D}'$ (so~$c_{2}^{e} + 1 < c_{1}^{e}$).
In any case, $\mathcal{D}^{e}$ has fewer crossings involving $e$ than $\mathcal{D}'$; a contradiction to $e$ creating a minimum number of additional crossings.
\end{proof}

\begin{figure}[htb]
\centering
\subcaptionbox{\label{fig:cross-min-extend-1}}[.49\textwidth]{\includegraphics[page=1]{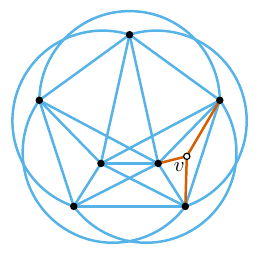}}
\subcaptionbox{\label{fig:cross-min-extend-2}}[.49\textwidth]{\includegraphics[page=2]{Figures/cross-min.pdf}}
\caption{\subref{fig:cross-min-extend-1}~A crossing-minimizing drawing on $8$ vertices that cannot be completed to a crossing-minimizing drawing of~$K_8$. The blue subdrawing is a crossing-minimizing drawing of~$K_7$. \subref{fig:cross-min-extend-2}~The shaded green areas and orange numbers indicate how many edges an edge starting at $v$ has to cross at least to reach the respective area or vertex.}
\label{fig:crossing-minimal}
\end{figure}

It is not too surprising that the resulting drawing $\mathcal{D}'$ need not be crossing-minimizing anymore (for example, when adding an edge between two isolated vertices lying in different faces of a crossing-free drawing).
\Cref{fig:cross-min-extend-1} shows a more involved example of a crossing-minimizing drawing~$\mathcal{D}$ on $8$ vertices that cannot be extended to a crossing-minimizing drawing of~$K_8$:
The subdrawing of $\mathcal{D}$ (drawn blue in \Cref{fig:cross-min-extend-1}) without $v$ is a drawing of $K_7$ with $9$~crossings, which is minimal.
There are four edges missing in $\mathcal{D}$, all of which are incident to~$v$.
The shaded green areas in \Cref{fig:cross-min-extend-2} indicate how many edges an edge starting in $v$ has to cross at least to reach that area.
Based on that, the numbers close to the vertices indicate how many edges an edge between $v$ and the respective vertex has to cross at least.
Summing up, we see that every extension of $\mathcal{D}$ to a simple drawing of $K_8$ has at least $19$~crossings.
This is one more than the minimum of $18$ crossings for drawings of~$K_8$.
Moreover, the way the example is constructed, it can be argued that every drawing weakly isomorphic to $\mathcal{D}$ is also strongly isomorphic to~$\mathcal{D}$.
Therefore, not even any drawing weakly isomorphic to $\mathcal{D}$ can be extended to a crossing-minimizing drawing of~$K_8$.

However, again similar to the proof for separable drawings, we can impose another minimality condition to add several edges at once.

\begin{mythm}\label{thm:extend-cross-min}
Let $\mathcal{D}$ be a crossing-minimizing drawing of a non-complete graph on $n$ vertices.
Then $\mathcal{D}$ can be extended to a simple drawing of~$K_n$.
\end{mythm}

\begin{proof}
We extend $\mathcal{D}$ to a drawing $\mathcal{D}'$ of $K_{n}$ such that (1) each added edge $e$ creates a minimum number of additional crossings in $\mathcal{D}$ and such that under this condition (2) $\mathcal{D}'$ has the least total number of crossings. Then, by \Cref{lem:extend-cross-min-by-one}, $\mathcal{D} + e$ is simple for each of those added edges.

Assume that $\mathcal{D}'$ is not simple. Then there are two added edges $e_{1}$ and $e_{2}$ in $\mathcal{D}'$ that have more than one point in common. Let $\mathcal{D}''$ be the drawing obtained from $\mathcal{D}'$ by exchanging $e_{1}$ and $e_{2}$ (between two consecutive common points on one of the edges). In the process, $e_{1}$ and $e_{2}$ exchange some crossings that they have with $\mathcal{D}$ and, by the first minimality condition on~$\mathcal{D}'$, the numbers of those exchanged crossings must coincide. Hence, also $\mathcal{D}''$ fulfills the first minimality condition but has in total at least one crossing less than $\mathcal{D}'$; a contradiction to the second minimality condition on $\mathcal{D}'$.
\end{proof}

\section{Crossing-Free Hamiltonian Cycles and Paths}\label{sec:hamiltonian}

This section is about separable drawings of the complete graph $K_n$.
We first show that they are plane Hamiltonian connected, that is, there exists a crossing-free Hamiltonian path between each pair of vertices, which proves \Cref{conj:paths} restricted to this class.
In parallel we also establish the basis to prove \Cref{conj:2n-3} restricted to separable drawings of $K_n$, that is, they contain a crossing-free subdrawing with at least $2n-3$ edges that in turn contains a Hamiltonian cycle.

\begin{mythm}\label{thm:paths-separable}
Every separable drawing $\mathcal{D}$ of $K_{n}$ with $n \geq 2$ vertices contains, for each pair of vertices $v \neq w$ in $\mathcal{D}$, a crossing-free subdrawing with at least $2n-3$ edges that in turn contains a Hamiltonian path with end-vertices $v$ and $w$.
\end{mythm}

\begin{figure}[htb]
\centering
\subcaptionbox{\label{fig:paths-separable}}[.32\textwidth]{\includegraphics[page=1]{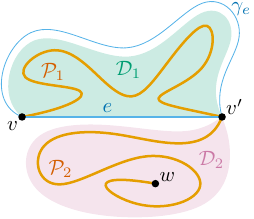}}
\subcaptionbox{\label{fig:cycle-separable}}[.32\textwidth]{\includegraphics[page=2]{Figures/hamiltonian.pdf}}
\subcaptionbox{\label{fig:edge-in-path-separable}}[.32\textwidth]{\includegraphics[page=3]{Figures/hamiltonian.pdf}}
\caption{Finding, in a separable drawing of $K_n$, a crossing-free \subref{fig:paths-separable}~Hamiltonian path between two given vertices, \subref{fig:cycle-separable}~Hamiltonian cycle, and \subref{fig:edge-in-path-separable}~Hamiltonian path containing a given edge $e$.}
\label{fig:hamiltonian-separable}
\end{figure}

\begin{proof}
The proof is by induction on $n$.
For $n = 2$ the drawing $\mathcal{D}$ consists of $2n-3 = 1$ edges and the statement is clearly true.
For the induction step, let $n \geq 3$, let $v \neq w$ be two arbitrary vertices in~$\mathcal{D}$, and consider some edge $e = \{ v, v' \}$ with $v' \neq w$ and witness~$\gamma_{e}$.
Further, let $\mathcal{D}_{1}$ be the subdrawing of $\mathcal{D}$ induced by the set of vertices on the side of $\gamma_{e}$ not containing $w$ and let $\mathcal{D}_{2}$ be the subdrawing of $\mathcal{D}$ induced by the set of vertices on the other side of $\gamma_{e}$ but without vertex~$v$.
See \Cref{fig:paths-separable} for an illustration.

Then $\mathcal{D}_{1}$ and $\mathcal{D}_{2}$ are both proper subdrawings of $\mathcal{D}$.
In particular, let $\mathcal{D}_{1}$ have $2 \leq k < n$ vertices and consequently $\mathcal{D}_{2}$ have $n-k+1$ vertices.
Hence, by the induction hypothesis, there exists a crossing-free subdrawing $\mathcal{D}_{1}'$ of $\mathcal{D}_{1}$ with at least $2k-3$ edges and a Hamiltonian path $\mathcal{P}_{1}$ in~$\mathcal{D}_{1}'$ with end-vertices $v$ and~$v'$.
Similarly there exists a crossing-free subdrawing $\mathcal{D}_{2}'$ of $\mathcal{D}_{2}$ with at least $2n-2k-1$ edges and a Hamiltonian path $\mathcal{P}_{2}$ in $\mathcal{D}_{2}'$ with end-vertices $v'$ and $w$.
By \Cref{lem:basic-separation}, the drawings $\mathcal{D}_{1}'$ and $\mathcal{D}_{2}'$ consist of disjoint sets of edges and, in particular, no edge of $\mathcal{D}_{1}'$ crosses any edge of $\mathcal{D}_{2}'$.
Consequently, the union of $\mathcal{D}_{1}'$ and $\mathcal{D}_{2}'$ forms a crossing-free subdrawing $\mathcal{D}'$ of~$\mathcal{D}$ with at least $2n-4$ edges.
Moreover, $v'$~is a cut vertex in $\mathcal{D}'$ and since maximal crossing-free subdrawings of simple drawings of $K_n$ are $2$-connected~\cite{gpt-2021-psctd}, we can find an edge $f$ in $\mathcal{D} \setminus \mathcal{D}'$ such that $\mathcal{D}' \cup f$ is still crossing-free and has at least $2n-3$ edges.
Furthermore, already $\mathcal{D}'$ contains the union of $\mathcal{P}_{1}$ and $\mathcal{P}_{2}$, which forms a crossing-free Hamiltonian path in $\mathcal{D}$ with end-vertices $v$ and~$w$.
This finishes the proof. 
\end{proof}

With a similar approach, now using \Cref{thm:paths-separable} instead of induction, we obtain that separable drawings of $K_n$ also contain a crossing-free subdrawing with at least $2n-3$ edges that in turn contains a Hamiltonian cycle, by this verifying the properties of \Cref{conj:cycles,conj:2n-3} for them.

\begin{mythm}\label{thm:cycle-separable}
Every separable drawing $\mathcal{D}$ of $K_{n}$ with $n \geq 3$ vertices contains a crossing-free subdrawing with at least $2n-3$ edges that in turn contains a Hamiltonian cycle.
\end{mythm}

\begin{proof}
Let $e = \{ v, w \}$ be an arbitrary edge in $\mathcal{D}$ with witness $\gamma_e$ and let $\mathcal{D}_{1}$ and $\mathcal{D}_{2}$ be the subdrawings of $\mathcal{D}$ induced by the vertices on the two sides of $\gamma_{e}$, respectively.
Let further $\mathcal{D}_{1}$ have $2 \leq k \leq n$ vertices and consequently $\mathcal{D}_{2}$ have $n-k+2$ vertices.
By \Cref{thm:paths-separable}, there exists a crossing-free subdrawing $\mathcal{D}_{1}'$ of $\mathcal{D}_{1}$ with at least $2k-3$ edges and a crossing-free subdrawing $\mathcal{D}_{2}'$ of~$\mathcal{D}_{2}$ with at least $2n-2k+1$ edges.
Moreover, $\mathcal{D}_{i}'$ contains a Hamiltonian path $\mathcal{P}_{i}$ with end-vertices $v$ and~$w$, for $i \in \{ 1, 2 \}$.
See \Cref{fig:cycle-separable} for an illustration.

By \Cref{lem:basic-separation} and similar as in the proof of \Cref{thm:paths-separable}, the union of $\mathcal{D}_{1}'$ and $\mathcal{D}_{2}'$ forms a crossing-free subdrawing $\mathcal{D}'$ of $\mathcal{D}$.
Since $e$ might be contained in both $\mathcal{D}_{1}'$ and $\mathcal{D}_{2}'$, the subdrawing $\mathcal{D}'$ has at least $2n-3$ edges.
Moreover, $\mathcal{D}'$ contains the union of $\mathcal{P}_{1}$ and $\mathcal{P}_{2}$, which forms a crossing-free Hamiltonian cycle in $\mathcal{D}$.
This finishes the proof. 
\end{proof}

Note in this context that in a g-convex drawing of $K_n$ actually every maximal plane subdrawing contains at least $2n-3$ edges~\cite{bfmos-2025-phccd}.
This does not hold for separable drawings because the example of Garc{\'i}a, Pilz, and Tejel \cite[Figure~4]{gpt-2021-psctd} with a maximal plane subdrawing with only $\frac{3n}{2}$ edges is a $2$-page book drawing and therefore separable.
However, as in \cite[Corollary~16]{bfmos-2025-phccd} for g-convex drawings, we can show that every edge $e$ in a separable drawing of~$K_n$ is part of some crossing-free Hamiltonian path.
Indeed, we can use the witness of $e$ to separate the drawing and then use \Cref{thm:paths-separable} on suitable subdrawings and end-vertices.
See \Cref{fig:edge-in-path-separable} for an illustration.

\begin{mycor}\label{cor:edge-in-hp-separable}
Every edge in a separable drawing $\mathcal{D}$ of $K_n$ is part of some crossing-free Hamiltonian path in~$\mathcal{D}$.
\end{mycor}

For the proofs of \Cref{thm:paths-separable,thm:cycle-separable} it is actually sufficient that for every pair of vertices $v$ and~$w$, one of them is incident to a separator edge that is not $\{ v, w \}$.
In particular, this is the case when every vertex is incident to at least $2$ separator edges.
We call this property \emph{degree-2-separable}.
In the proof we further rely on induction.
Therefore, we call a class $\mathcal{S}$ of simple drawings \emph{subset-closed} if every subdrawing of a drawing in $\mathcal{S}$ is itself in $\mathcal{S}$.
With this we get the following observation, which might be helpful to show the properties of \Cref{conj:cycles,conj:paths,conj:2n-3} for even larger classes of simple drawings.

\begin{myobs}\label{obs:deg-2-separable}
Let $\mathcal{S}$ be a subset-closed class of simple drawings of complete graphs such that every drawing in $\mathcal{S}$ is degree-2-separable.
Then every drawing in $\mathcal{S}$ contains a crossing-free Hamiltonian cycle.
\end{myobs}

Let us further mention that a single separator edge is enough to find a crossing-free matching of linear size; let us call this property \emph{1-separable} for a subset-closed class of simple drawings.
Indeed, we can add the separator edge $e$ to the matching and then recurse on the subdrawings in the two sides of the witness~$\gamma_e$.
In the worst case, for each edge that we add, we get two subdrawings with only one vertex each that cannot be matched anymore.

\begin{myobs}
Let $\mathcal{S}$ be a subset-closed class of simple drawings of complete graphs such that every drawing in $\mathcal{S}$ is 1-separable.
Then every drawing in $\mathcal{S}$ contains a crossing-free matching of size linear in the number of its vertices.
\end{myobs}

\begin{figure}[htb]
\centering
\subcaptionbox{\label{fig:non-separable-1}}[.49\textwidth]{\includegraphics[page=1]{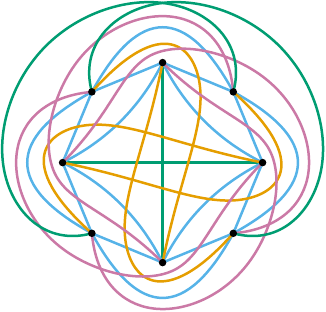}}
\subcaptionbox{\label{fig:non-separable-2}}[.49\textwidth]{\includegraphics[page=2]{Figures/non-separable.pdf}}
\caption{The only two simple drawings of $K_8$ that do not have a single separator edge.}
\label{fig:non-separable}
\end{figure}

Unfortunately there exist simple drawings of $K_n$ without a single separator edge.
\Cref{fig:non-separable} shows the (up to weak isomorphism) only two simple drawings of $K_8$ with this property; the different edge colors are just for better visibility.
We obtained these drawings by applying the algorithm of \Cref{thm:poly} (see \Cref{sec:recognition} below) to all different rotation systems of $K_8$~\cite{aafhpprsv-2015-agdscg}.
Note that Harborth and Mengersen~\cite{hm-1974-ewcdcg} proved that simple drawings of $K_n$ for $n \leq 7$ always have uncrossed edges, and therefore, they have separator edges by \Cref{obs:uncrossed}.
Hence, the drawings depicted in \Cref{fig:non-separable} are the smallest examples without any separator edge.

We continue this section by proving that all g-convex drawings are separable, therefore showing that our results on plane Hamiltonicity improve upon the work of Bergold, Felsner, M.~Reddy, Ort\-haber, and Scheucher~\cite{bfmos-2025-phccd}.
Our proof is inspired by the proof of Arroyo, Richter, and Sunohara~\cite{ars-2021-edcgap} that all so-called hereditarily convex drawings (of $K_n$) are pseudospherical.

\begin{mythm}\label{thm:gconvex}
Every g-convex drawing (of $K_n)$ is separable.
\end{mythm}

\begin{proof}
We show that every edge $e = \{ a, b \}$ in a g-convex drawing $\mathcal{D}$ is a separator edge.
If $e$ is uncrossed, then it is a separator edge by \Cref{obs:uncrossed}.
Hence, we can assume that $e$ is crossed by at least one edge.
In the following we find a simple closed curve $\gamma_e$ fulfilling the weaker condition in \Cref{lem:separator-easier-kn}, thereby showing that $e$ is a separator edge.
In particular, we find vertex sets $V_L$ and $V_R$ that will correspond to the vertices on the two sides of $\gamma_e$, respectively.

We fix an orientation of $e$ and say that a vertex $v$ of $\mathcal{D}$ lies on \emph{the left} or on \emph{the right} of $e$ if the convex side of the triangle spanned by $e$ and $v$ lies to the left or right of the oriented edge $e$, respectively.
Recall that both sides of such a triangle can be convex and that the convex side is unique if and only if it is part of a crossing $K_4 = \{ a, b, v, w \}$.
In this~$K_4$, $e$~can either be a diagonal or a boundary edge.
In the first case, $v$ and $w$ lie on different sides of $e$ and, in the second case, they lie on the same side of $e$; see \Cref{fig:4-tuple} for an illustration.

\begin{figure}[htb]
\centering
\subcaptionbox{\label{fig:4-tuple-1}}[.40\textwidth]{\includegraphics[page=3]{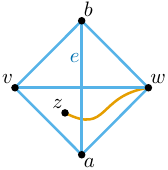}}
\subcaptionbox{\label{fig:4-tuple-2}}[.40\textwidth]{\includegraphics[page=4]{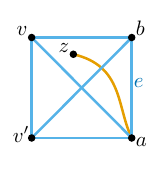}}
\caption{In a crossing $K_4 = \{ a, b, v, w \}$, the edge $e$ is either \subref{fig:4-tuple-1}~a diagonal edge or \subref{fig:4-tuple-2}~a boundary edge. In our algorithm, every vertex $z$ that lies in the convex side of a triangle spanned by $e$ and some $v \in V_L$ is also added to~$V_L$.}
\label{fig:4-tuple}
\end{figure}

We start with $V_L = V_R = \emptyset$.
In a first step, we consider crossing $K_4$'s where $e$ is a diagonal and we add the respective vertices $v$ that are to the left of $e$ to~$V_L$.
Since $\mathcal{D}$ is g-convex, the respective vertices $w$ that are to the right of $e$ will never be added to~$V_L$.
In a second step, we successively add vertices $v$ to $V_L$ if there exists a crossing $K_{4} = \{ a, b, v', v \}$ such that $e$ is a boundary edge and $v'$ was already added to $V_L$ before.
Once we cannot add anymore vertices to $V_L$ in this manner, we add all remaining vertices to~$V_R$.
Note that for all vertices $v$ in $V_L$ the unique convex side of the triangle spanned by $v$ and $e$ is to the left of~$e$.

Let $\mathcal{D}_{L}$ and $\mathcal{D}_{R}$ be the subdrawings of $\mathcal{D}$ induced by $e$ and the vertices $V_L$ and $V_R$, respectively.
Note that $e$ is uncrossed in both those subdrawings.
Consider the cell~$F_{\infty}$ in~$\mathcal{D}_L$ that is incident to $e$ and to its right.
We show that all vertices of $V_R$ lie in~$F_{\infty}$.

Towards a contradiction, assume first that a vertex $z$ of $V_R$ lies in the convex side of a triangle spanned by $e$ and a vertex $v$ added to $V_L$ in the first step, that is, there is an edge $\{ v, w \}$ that crosses~$e$.
Then, by convexity, the edge $\{ z, w \}$ lies in the crossing side of the $K_4$ spanned by $\{ a, b, v, w \}$ and hence $\{ z, w \}$ also crosses $e$; see \Cref{fig:4-tuple-1} for an example.
This implies that $z$ lies to the left of $e$ and was added to $V_L$ in the first step; a contradiction to~$z \in V_R$.

Assume next that $z$ lies in the unique convex side of a triangle spanned by $e$ and a vertex $v$ added to $V_L$ in the second step.
In particular, let $v$ be the first vertex added to $V_L$ such that the unique convex side of $\{ a, b, v \}$ contains~$z$.
Then either $\{ z, a \}$ or $\{ z, b \}$ crosses a triangle spanned by $e$ and some vertex $v'$ added to $V_L$ before~$v$; see \Cref{fig:4-tuple-2} for an example.
Therefore, $z$ is added to $V_L$ in the second step; again a contradiction.

Assume last that $z$ lies neither in $F_{\infty}$ nor in the convex side of any of the triangles spanned by $e$ and~$V_L$.
Then the edges $\{ z, a \}$ and $\{ z, b \}$ cannot cross any of those triangle edges because otherwise $z$ would have been added to $V_L$ in the second step.
Further, no vertex of $V_L$ can lie in the triangle spanned by $z$ and $e$ to the right of $e$ because we would not have added it to $V_L$ then.
Since $z$ does not lie in $F_{\infty}$, some part of $\mathcal{D}_{L}$  separates it from there.
Hence, there is an edge $f = \{ v_{1}, v_{2} \}$ in $\mathcal{D}_{L}$ crossing $\{ z, a \}$ or $\{ z, b \}$.
Since $v_{1}$ and $v_{2}$ are on the same side of the triangle $\{ z, a, b \}$ and $f$ does not cross~$e$, it follows that $f$ crosses both $\{ z, a \}$ and $\{ z, b \}$; see \Cref{fig:5-tuple-1} for an illustration.
Consequently, the triangles $\{ v_{1}, v_{2}, a \}$ and $\{ v_{1}, v_{2}, b \}$ have no convex side; a contradiction to $\mathcal{D}$ being g-convex.
This finished the proof that all vertices of $V_R$ lie in~$F_{\infty}$.

\begin{figure}[htb]
\centering
\subcaptionbox{\label{fig:5-tuple-1}}[.40\textwidth]{\includegraphics[page=5]{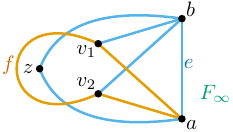}}
\subcaptionbox{\label{fig:5-tuple-2}}[.40\textwidth]{\includegraphics[page=6]{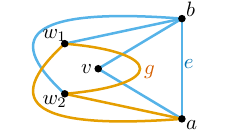}}
\caption{\subref{fig:5-tuple-1}~A vertex $z \in V_R$ that lies neither in $F_{\infty}$ nor in the convex side of any of the triangles spanned by $e$ and~$V_L$. \subref{fig:5-tuple-2}~An edge $g$ of $\mathcal{D}_{R}$ that crosses an edge of $\mathcal{D}_{L}$ incident to~$e$. Both situations lead to a triangle (marked orange) not having a convex side.}
\label{fig:5-tuple}
\end{figure}

It remains to show that no edge $g = \{ w_{1}, w_{2} \}$ of $\mathcal{D}_{R}$ can cross any edge of~$\mathcal{D}_{L}$.
Towards a contradiction, assume first that $g$ crosses an edge incident to~$e$ and some vertex $v$ of~$V_L$.
If $g$ is also incident to $e$, without loss of generality $g = \{ w_1, a \}$, then we would have added $w_1$ to $V_L$ in the second step; a contradiction.
Hence, both $w_1$ and $w_2$ are in~$F_\infty$ and $g$ crosses both $\{ v, a \}$ and~$\{ v, b \}$.
Further, by the argument that $g$ cannot be incident to~$e$, no edge between $w_1$ or $w_2$ and $a$ or $b$ crosses $\{ v, a \}$ or~$\{ v, b \}$.
This leaves only one possibility, up to strong isomorphism, to realize the situation as a simple drawing; see \Cref{fig:5-tuple-2} for an illustration.
As a result, the triangles $\{ w_{1}, w_{2}, a \}$ and $\{ w_{1}, w_{2}, b \}$ have no convex side; a contradiction.
Assume second that $g$ crosses an edge $f$ in $\mathcal{D}_{L}$ that is independent from~$e$ but $g$ crosses no edge incident to~$e$ in $\mathcal{D}_{L}$.
Since at least one end-vertex of $g$ lies in $F_{\infty}$, this is only possible if $f = \{ v, v' \}$ as in \Cref{fig:4-tuple-2} and $g$ crosses $f$ at least twice; a contradiction to being a simple~drawing.

Hence, we can close the edge $e$ in $F_\infty$ between the boundaries of $\mathcal{D}_L$ and $\mathcal{D}_R$ to a simple curve $\gamma_e$ that fulfills all properties of a witness, that is, $e$ is a separator edge.
\end{proof}

\begin{figure}[htb]
\centering
\subcaptionbox{\label{fig:g-convex-left-right}}[.49\textwidth]{\includegraphics[page=1]{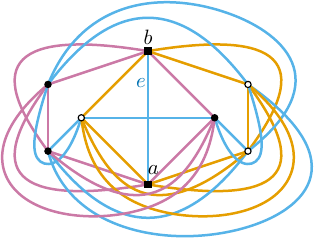}}
\subcaptionbox{\label{fig:non-g-convex-or-2-page}}[.49\textwidth]{\includegraphics[page=2]{Figures/g-convex.pdf}}
\caption{\subref{fig:g-convex-left-right}~A g-convex drawing where a complete left-right splitting via convex sides is not possible: The subdrawing induced by $e$ and all vertices on the left of $e$ (marked with circles) is drawn orange, the respective subdrawing on the right of $e$ is drawn purple. \subref{fig:non-g-convex-or-2-page}~A separable drawing that is neither g-convex nor a 2-page book drawing.}
\label{fig:g-convex-example}
\end{figure}

Note that we could not just add \emph{all} vertices to the left or right of $e$ to $V_L$ or $V_R$, respectively.
\Cref{fig:g-convex-left-right} shows an example where this would not result in two separated subdrawings.

To see that separable drawings are not only the union of g-convex and 2-page book drawings, for example, consider an arbitrary straight-line drawing with $k \geq 5$ vertices $\{ v_1, \ldots, v_k \}$ on the convex hull and reroute the edges $\{ v_1, v_3 \}$ and $\{ v_2, v_4 \}$ outside of the convex hull.
The resulting drawing is always separable (each edge has a witnesses that is straight-line within the convex hull and closed on the outside), not g-convex (the triangles $\{ v_1, v_2, v_4 \}$ and $\{ v_1, v_3, v_4 \}$ have no convex side), and in most cases also not weakly isomorphic to any 2-page book drawing. \Cref{fig:non-g-convex-or-2-page} shows the smallest such example.

\begin{figure}[htb]
\centering
\subcaptionbox{\label{fig:non-compatible-star-cfhc-8}}[.32\textwidth]{\includegraphics[page=1]{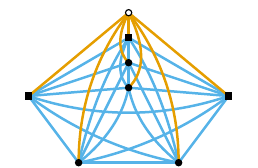}}
\subcaptionbox{\label{fig:non-compatible-star-cfhc-9}}[.32\textwidth]{\includegraphics[page=2]{Figures/non-compatible-star-cycle.pdf}}
\subcaptionbox{\label{fig:non-compatible-star-cfhc-n-4}}[.32\textwidth]{\includegraphics[page=3]{Figures/non-compatible-star-cycle.pdf}}
\caption{\subref{fig:non-compatible-star-cfhc-8}~The only separable drawing of $K_8$ containing a spanning star (drawn orange) that is not compatible with any crossing-free Hamiltonian cycle. \subref{fig:non-compatible-star-cfhc-9}~The only separable drawing of $K_9$ containing two such spanning stars (centered at the vertices marked with circles). \subref{fig:non-compatible-star-cfhc-n-4}~A separable drawing of $K_{20}$ such that the spanning stars of the $5$ vertices on the convex hull (marked with circles) are not compatible with any crossing-free Hamiltonian cycle. The only non-straight-line edges are indicated in orange.}
\label{fig:non-compatible-star-cfhc}
\end{figure}

Moreover, recall that the main structural result in~\cite{bfmos-2025-phccd} states that every spanning star in a g-convex drawing is compatible with some crossing-free Hamiltonian cycle.
That is, for every spanning star $\mathcal{S}$ there exists a crossing-free Hamiltonian cycle $\mathcal{C}$ such that $\mathcal{S} \cup \mathcal{C}$ is still crossing-free.
This clearly also holds for $2$-page book drawings of $K_n$ since they contain a completely uncrossed Hamiltonian cycle.
However, we found a separable drawing $\mathcal{D}_8$ of $K_8$ with a spanning star that is not compatible with any crossing-free Hamiltonian cycle.
See \Cref{fig:non-compatible-star-cfhc-8} for a depiction; the spanning star is marked in orange.
$\mathcal{D}_8$ is a straight-line drawing except for the subdrawing induced by the $4$ vertices in the middle, which form a $2$-page book drawing.
This makes it easy to verify that it is a separable drawing.
Moreover, note that the potential union of a spanning star $\mathcal{S}$ with a crossing-free Hamiltonian cycle can also be seen as an edge-disjoint union of $\mathcal{S}$ with a crossing-free Hamiltonian path $\mathcal{P}$ on the remaining $n-1$ vertices.
And $\mathcal{D}_8$ is constructed in such a way that at least $3$ vertices (marked with squares) would have to be end-vertices of $\mathcal{P}$.

This constitutes a clear difference between separable drawings and the union of g-convex and $2$-page book drawings.
We also found a separable drawing $\mathcal{D}_9$ of $K_9$ that contains two spanning stars which are not compatible with any crossing-free Hamiltonian cycle.
See \Cref{fig:non-compatible-star-cfhc-9} for a depiction.
The argumentation for non-compatibility is similar as before and symmetric for the spanning stars of the two vertices marked with circles.
Further, the union of all completely uncrossed edges in $\mathcal{D}_9$ is $2$-connected, which can be shown to always lead to a separable drawing.
Hence, these two drawings also give additional ideas on how to construct separable drawings that are neither g-convex nor $2$-page book drawings.
Furthermore, based on computations on the rotation system database~\cite{aafhpprsv-2015-agdscg}, $\mathcal{D}_8$ is the only separable drawing of~$K_8$ with a spanning star that is not compatible with any crossing-free Hamiltonian cycle, and $\mathcal{D}_9$ is the only separable drawing of $K_9$ with more than one such spanning star.
Based on~$\mathcal{D}_8$, we can however construct separable drawings of $K_n$ were $\lfloor n/4 \rfloor$ of its spanning stars are not compatible with any crossing-free Hamiltonian cycle.
\Cref{fig:non-compatible-star-cfhc-n-4} gives an example illustration.

\begin{myobs}\label{obs:separable-star-no-compatible-hc}
There exist separable drawings of $K_n$ with $n \geq 20$ vertices such that at least $\lfloor n/4 \rfloor$ of its spanning stars are not compatible with any crossing-free Hamiltonian cycle.
\end{myobs}

\section{Flips in Rotation Systems}\label{sec:flips}

In this section we make a connection between separable drawings of $K_n$ and their rotation systems.
Specifically, we show that for a simple drawing $\mathcal{D}$ of~$K_n$, separability only depends on the rotation system of~$\mathcal{D}$.
To this end, we first introduce local changes in rotation systems, which we call flips.

For the following discussion, we denote an abstract rotation system as \emph{realizable rotation system} if it is the rotation system of a simple drawing;
otherwise we denote it as \emph{non-realizable rotation system}.

\begin{figure}[htb]
\centering
\subcaptionbox{\label{fig:flip-rs}}[.32\textwidth]{\includegraphics[page=1]{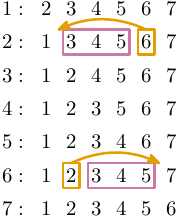}}
\subcaptionbox{\label{fig:flip-drawing}}[.32\textwidth]{\includegraphics[page=2]{Figures/flip.pdf}}
\subcaptionbox{\label{fig:non-realizable-flip}}[.32\textwidth]{\includegraphics[page=3]{Figures/flip.pdf}}
\caption{\subref{fig:flip-rs}~A rotation system corresponding to a convex straight-line drawing of $K_7$. The only (potential) flip of the edge $e = \{ 2, 6 \}$ is marked. \subref{fig:flip-drawing}~A corresponding drawing with the flip of $e$ marked in orange (the solid version is before the flip, the dashed version after). \subref{fig:non-realizable-flip}~A potential flip in a simple drawing of $K_5$ where the resulting rotation system is not realizable.}
\label{fig:flip}
\end{figure}

A \emph{potential flip} of an edge $e = \{ v, w \}$ in a realizable rotation system of $K_n$ is the following operation (see \Cref{fig:flip-rs} for an example illustration):
\begin{itemize}
\item Remove $e$ in the rotations of $v$ and $w$, and add it again at different positions such that, in the counter-clockwise rotation of $v$ and the clockwise rotation of $w$, the sets of vertices between the position of $e$ before and after the operation coincide and are non-empty.
\end{itemize}
If the resulting (abstract) rotation system after a potential flip is still realizable, then we call the operation a \emph{flip}.
As we show in \Cref{thm:flip-in-rs-and-drawing}, a flip of $e$ in the rotation system corresponds to redrawing $e$ in an arbitrary simple drawing realizing the rotation system.
See \Cref{fig:flip-drawing} for an example drawing corresponding to the flip in \Cref{fig:flip-rs}.

Already for realizable rotation systems of $K_5$ it is possible that a potential flip leads to a non-realizable rotation system.
\Cref{fig:non-realizable-flip} shows a drawing corresponding to such a situation; in the drawing the flipped edge (dashed orange) would cross the purple edge twice.
However, as we show next, a potential flip keeps all $4$-tuples in the rotation system realizable.

\begin{mylem}\label{lem:flip-4-tuples}
After a potential flip of an edge $e = \{ v, w \}$ in a realizable rotation system of~$K_n$ with $n \geq 4$ all $4$-tuples of vertices in the new rotation system are still realizable.
\end{mylem}

\begin{figure}[htb]
\centering
\subcaptionbox{\label{fig:flip-k4-base}}[.32\textwidth]{\includegraphics[page=4]{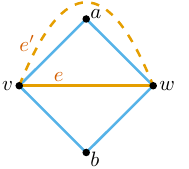}}
\subcaptionbox{\label{fig:flip-k4-crossing}}[.32\textwidth]{\includegraphics[page=5]{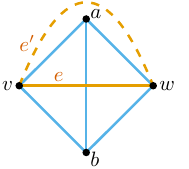}}
\subcaptionbox{\label{fig:flip-k4-uncrossed}}[.32\textwidth]{\includegraphics[page=6]{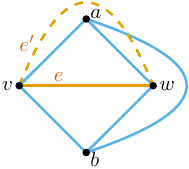}}
\caption{\subref{fig:flip-k4-base}~The only basic configuration how a flip of $e$ can change the rotation system induced by $4$ vertices. \subref{fig:flip-k4-crossing}~and \subref{fig:flip-k4-uncrossed}~show the two cases how the edge $\{ a, b \}$ can be added to this configuration.}
\label{fig:flip-k4}
\end{figure}

\begin{proof}
When flipping $e$ to~$e'$, the induced rotation system on $4$ vertices can only be affected if both $v$ and $w$ are part of that $4$-tuple.
Hence, consider a $4$-tuple $V = \{ v, w, a, b \}$.
If the flip changes the rotation system induced by $V$, then $a$ lies in the counter-clockwise rotation of $v$ and the clockwise rotation of $w$ between the position of $e$ and~$e'$, and $b$ lies outside of these intervals.
See \Cref{fig:flip-k4-base} for an illustration.
This fixes all rotations on that $4$-tuple, except for the edge~$\{ a, b \}$.
Since the rotation system is realizable before the flip, there are exactly two possibilities, how $\{ a, b \}$ can be added.
One corresponds to a drawing of $K_4$ where $e$ crosses $\{ a, b \}$ (\Cref{fig:flip-k4-crossing}), the other corresponds to a drawing of $K_4$ without crossings (\Cref{fig:flip-k4-uncrossed}).
After the flip, in the first case the rotation system corresponds to a drawing of $K_4$ without crossings, in the second case it corresponds to a drawing of $K_4$ where $e'$ crosses $\{ a, b \}$.
In particular, the rotation system induced by $V$ after the flip is still realizable, which finishes the proof.
\end{proof}

Before we come to the main result of this section, recall that an abstract rotation system of $K_n$ is realizable if and only if all induced rotation systems on subsets of $5$ vertices are realizable.
When flipping an edge $e = \{ v, w \}$, only rotation systems induced by $5$ vertices that involve both $v$ and $w$ can change, and we have to test them for realizability.
There are $\binom{n-2}{5-2}=\Theta(n^3)$ such sets and each can be tested in constant time.

\begin{myobs}\label{obs:test-realizable-flip}
It can be tested in $\mathcal{O}(n^3)$ time whether a realizable rotation system of~$K_n$ is still realizable after a potential flip.
\end{myobs}

To relate separator edges to flips in rotation systems we make use of a result by Schaefer \cite[Corallary~3.7]{s-2021-tdgtpdcg}, which is a generalization of Gioan's Theorem~\cite{g-2022-cgdtm}.
It states that every pair of drawings of $K_n$ minus a non-perfect matching having the same set of crossings can be transformed into each other via triangle mutations (plus a homeomorphism of the sphere).
A \emph{triangle mutation} is the operation of moving an edge over the crossing between two other edges; see also \Cref{fig:triangle-mutation}.
It is important to note that Schaefer's result only holds for drawings and not for rotation systems.

\begin{mythm}\label{thm:flip-in-rs-and-drawing}
A flip of an edge $e$ in a realizable rotation system is equivalent to redrawing~$e$ in an arbitrary realization $\mathcal{D}$ of that rotation system such that the resulting drawing has a different rotation system and is still simple.
\end{mythm}

\begin{proof}
For the first direction, let $e$ be flipped to $e'$ and let $\mathcal{D}'$ be an arbitrary simple drawing realizing the rotation system with $e'$ instead of $e$.
Recall that for a rotation system of $K_n$, $n \geq 4$, the subrotation systems induced by the $4$-tuples of vertices determine all crossings and that there can be at most one crossing per $4$-tuple.
Following the proof of \Cref{lem:flip-4-tuples}, when such an induced rotation system is changed by the flip, then either $e$ or $e'$ is involved in a crossing on that $4$-tuple.
This implies that $\mathcal{D} - e$ and $\mathcal{D}' - e'$ have the same crossing edge pairs, that is, they at most differ in the order of crossings along edges.
Hence we can apply Schaefer's generalization of Gioan's theorem to transform $\mathcal{D}' - e'$ to $\mathcal{D}- e$ via triangle mutations.
More precisely, we apply those triangle mutations to transform $\mathcal{D}'$ and, whenever we would move an edge $f$ over a crossing and $e'$ lies between $f$ and that crossing, we first move $e'$ over the crossing and then make the originally planned move with $f$; see \Cref{fig:triangle-mutation} for an illustration.
This process may change the order of crossings along~$e'$, but changes neither the crossing edge pairs nor the rotation at any vertex.
Hence, once we have transformed $\mathcal{D}'- e' $ to $\mathcal{D}- e$, we have obtained a transformed edge $e'$ in the process such that $\mathcal{D} - e + e'$ is a realization of the flipped rotation system.
This shows that the flip of $e$ corresponds to redrawing $e$ by the transformed edge $e'$ in an arbitrary realization $\mathcal{D}$ of the original rotation system, which finishes the first direction.

\begin{figure}[htb]
\centering
\subcaptionbox{\label{fig:mutation-e-between}}[.32\textwidth]{\includegraphics[page=1]{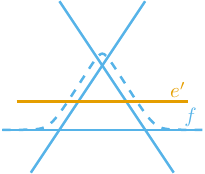}}
\subcaptionbox{\label{fig:mutation-first-e}}[.32\textwidth]{\includegraphics[page=2]{Figures/triangle-mutation.pdf}}
\subcaptionbox{\label{fig:mutation-then-f}}[.32\textwidth]{\includegraphics[page=3]{Figures/triangle-mutation.pdf}}
\caption{\subref{fig:mutation-e-between}~If after a triangle mutation the redrawn edge $f$ (dashed) would cross $e'$ twice, then \subref{fig:mutation-first-e}~we first move $e'$ over the respective crossing and then \subref{fig:mutation-then-f}~redraw $f$ as planned.}
\label{fig:triangle-mutation}
\end{figure}

For the other direction, let $e = \{ v, w \}$ be redrawn by $e'$ in $\mathcal{D}$ and let $\mathcal{D}'$ be the resulting simple drawing.
Since $\mathcal{D}$ and $\mathcal{D}'$ have different rotation systems, the rotation of $v$ and/or~$w$ has changed by replacing $e$ with~$e'$; without loss of generality, let the rotation of~$v$ be different.
Further, since both $\mathcal{D}$ and $\mathcal{D}'$ are simple drawings, neither $e$ nor $e'$ crosses any edge $\{ v, a \}$ or $\{ a, w \}$.
Therefore, every vertex $a$ in the counter-clockwise rotation of $v$ between the position of $e$ and $e'$ lies between the position of $e$ and $e'$ in the clockwise rotation of~$w$.
This is exactly the definition of a flip in the rotation system, which finishes the proof.
\end{proof}

Based on \Cref{thm:flip-in-rs-and-drawing}, we can use flips interchangeably for simple drawings and their rotation systems.
This gives, as we show next, an alternative definition of a separator edge for simple drawings of $K_n$ that is based on flips in rotation systems.
In particular, it makes being separable a property of the rotation system for simple drawings of~$K_n$.

\begin{mythm}\label{thm:separator-flip-equivalence}
Let $\mathcal{D}$ be a simple drawing of $K_{n}$ and let $e = \{ v, w \}$ be an edge of~$\mathcal{D}$.
Then $e$ is a separator edge if and only if $e$ is either uncrossed or can be flipped to an edge $e'$ such that $e$ and $e'$ cross disjoint sets of edges.
\end{mythm}

\begin{proof}
For the first direction, assume that $e$ is a separator edge and recall that $\gamma_{e}$ has at most one point in common with every edge $f \neq e$ in $\mathcal{D}$.
Hence, replacing $e$ in $\mathcal{D}$ by $e' = \gamma_e \setminus e$ gives a simple drawing $\mathcal{D}'$.
If $\mathcal{D}$ and~$\mathcal{D}'$ have the same crossings, then both $e$ and $e'$ are uncrossed.
Otherwise the rotation system has changed and, by \Cref{thm:flip-in-rs-and-drawing}, this corresponds to a flip of~$e$.

For the other direction, if $e$ is uncrossed, then $e$ is a separator edge by \Cref{obs:uncrossed}.
So assume that $e$ can be flipped to $e'$ such that no edge is crossed by both $e$ and~$e'$.
By \Cref{thm:flip-in-rs-and-drawing} we can redraw $e$ as $e'$ in~$\mathcal{D}$.
Further, there is some $4$-cycle $(v, a, w, b)$ in $\mathcal{D}$ separating $e$ from $e'$; see also \Cref{fig:flip-k4}.
Hence, the union $e \cup e'$ forms a simple closed curve~$\gamma_e$.
And since $e$ and $e'$ cross disjoint sets of edges, $\gamma_e$ crosses every edge of $\mathcal{D}$ at most once.
In other words, $\gamma_e$ is a witness for~$e$.
\end{proof}

\begin{figure}[htb]
\centering
\subcaptionbox{\label{fig:flippable-not-separator}}[.24\textwidth]{\includegraphics[page=1]{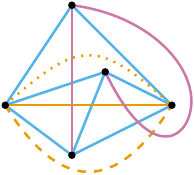}}
\subcaptionbox{\label{fig:non-witness-flip}}[.24\textwidth]{\includegraphics[page=2]{Figures/flip-vs-separator.pdf}}
\subcaptionbox{\label{fig:witness-flip}}[.24\textwidth]{\includegraphics[page=3]{Figures/flip-vs-separator.pdf}}
\subcaptionbox{\label{fig:subset-not-flippable}}[.25\textwidth]{\includegraphics[page=4]{Figures/flip-vs-separator.pdf}}
\caption{\subref{fig:flippable-not-separator}~The orange edge has two possible flips (dashed/dotted) but none of those yields a witness (one of the two purple edges would be crossed twice). \subref{fig:non-witness-flip}~A flip in a separable drawing (orange edge from solid to dashed) that leads to a non-separable drawing. This can even happen \subref{fig:witness-flip}~when flipping along a witness (the purple edge is not a separator edge anymore after flipping the orange edge). \subref{fig:subset-not-flippable}~When removing the gray vertex the orange edge is not flippable anymore.}
\label{fig:flip-vs-separator}
\end{figure}

Let us conclude this section with some remarks:
There are two basic differences between separator edges and edges that can be flipped.
First, every uncrossed edge is a separator edge but uncrossed edges are not necessarily flippable.
This is the case, for example, for edges on the convex hull in the convex straight-line drawing of~$K_n$.
Second, if a flipped edge~$e'$ crosses an edge~$f$ that was also crossed by the original edge~$e$, then $e$ is not a separator edge (unless there exists another suitable flip for~$e$).
\Cref{fig:flippable-not-separator} shows an example in the twisted drawing of~$K_5$, which is (up to strong isomorphism) the only drawing of $K_5$ that is not separable.
When making such a flip in a separable drawing, note that the resulting drawing might not be separable anymore.
See \Cref{fig:non-witness-flip} for an example in a separable drawing of~$K_5$ (which incidentally is not g-convex).
Furthermore, even flipping an edge $e$ along a witness $\gamma_e$, that is, replacing $e$ by $e' = \gamma_e \setminus e$ in a separable drawing (which makes sure that $e'$ is still a separator edge) might lead to a drawing that is not separable anymore.
This is because the witness $\gamma_f$ of some other edge might have to cross $e'$ twice; \Cref{fig:witness-flip} shows such an example with unique witnesses (up to strong isomorphism).
Last but not least, a separator edge in a simple drawing $\mathcal{D}$ is still a separator edge in every subdrawing of~$\mathcal{D}$.
Contrarily, an edge $e$ that can be flipped in a simple drawing of $K_n$ might not be flippable anymore in a vertex-induced subdrawing because we can remove all the vertices that $e$ was flipped over.
In other words, being a separator edge is a subset-closed property while being flippable is not.
An easy example for the latter is that every simple drawing of $K_4$ has a flippable edge but no edge in a simple drawing of $K_3$ is flippable.
One might argue that we should make an exception for edges that become uncrossed.
However, \Cref{fig:subset-not-flippable} shows a more involved example where the edge $e$ (orange) is still crossed but unflippable after removing a vertex (clearly in this case $e$ also cannot be a separator edge though).

\section{Recognition}\label{sec:recognition}

Using flips in rotation systems, especially \Cref{thm:separator-flip-equivalence}, we obtain a polynomial time recognition algorithm for separable drawings of~$K_n$.
To ease argumentation, we state the theorem in terms of rotation systems.
If the drawing is given differently, the time for calculating the rotation system has to be added, which is, however, dominated by the overall running time for all representations of a drawing that we know of.

\begin{mythm}\label{thm:poly}
It can be decided in $\mathcal{O}(n^6)$ time whether a simple drawing $\mathcal{D}$ of $K_{n}$, given by its rotation system, is separable.
\end{mythm}

\begin{proof}
We check, for each edge $e = \{ v, w \}$ in $\mathcal{D}$, whether it is a separator edge.
If $e$ is uncrossed, then it is a separator edge by \Cref{obs:uncrossed} and we are done.
Otherwise we use the relation between separator edges and flips in rotation systems given by \Cref{thm:separator-flip-equivalence}.
In particular, we determine all possible flips of $e$ and check whether one of them yields a witness for~$e$.

By the definition of a flip of $e$ in the rotation system, the subsets in the counter-clockwise rotation of $v$ and the clockwise rotation of $w$ between the position of $e$ before and after the flip coincide.
We get all such possibilities of potential flips for $e$ in $\mathcal{O}(n)$ time, by going through the rotations of $v$ and $w$ in parallel, starting with $\{ v, w \}$.
In this process we keep a parity list of all vertices and how often they appeared in the subsets.
Further, we use a counter to see how many of the vertices appeared an odd number of times, that is, showed up in only one of the two subsets so far.
Every time this counter is zero we have a potential flip.

By \Cref{obs:test-realizable-flip}, checking whether the new rotation system after a potential flip is realizable takes $\mathcal{O}(n^3)$ time.
To check whether the flipped edge has all different crossings from the original edge, we test for all $\mathcal{O}(n^2)$ new crossings whether they also existed before.
Since each crossing can be checked in constant time, this takes $\mathcal{O}(n^2)$ time for all crossings, which is less than the $\mathcal{O}(n^3)$ time for checking realizability.

Overall there are $\mathcal{O}(n^2)$ many edges $e$, each of them has $\mathcal{O}(n)$ potential flips, and testing whether such a flip yields a witness for $e$ takes $\mathcal{O}(n^3)$ time as argued.
Hence, we can decide in $\mathcal{O}(n^6)$ time whether a simple drawing of $K_{n}$ is separable.
\end{proof}

Unfortunately, the situation is very different for simple drawings of arbitrary graphs.
In particular, in the following we construct simple drawings of matchings where it is NP-hard to decide whether they are separable.
For this we use a reduction from linked planar 3-SAT with negated edges on one side, which was shown to be NP-hard by Pilz~\cite[Theorem~10]{p-2019-p3satcvc}.

The incidence graph $G_\phi$ of a 3-SAT formula $\phi$ has one vertex for each variable and each clause in $\phi$ and an edge between a variable vertex and a clause vertex if the variable occurs in the clause (as a positive or negative literal). If $G_\phi$ is a planar graph, then $\phi$ is a \emph{planar} 3-SAT instance.
For \emph{linked} planar 3-SAT there is a Hamiltonian cycle $\mathcal{C}$ that first visits all variable vertices and then all clause vertices such that the union of $G_\phi$ and $\mathcal{C}$ is still a planar graph.
Further, in the restriction \enquote{with negated edges on one side}, there exists an embedding of $G_\phi \cup \mathcal{C}$ such that all edges in $G_\phi$ corresponding to positive literals are drawn inside of $\mathcal{C}$ and all edges corresponding to negative literals are drawn outside of~$\mathcal{C}$.
(Note that a clause can still contain positive and negative literals.)

The main reason why we need this involved version of planar 3-SAT is to make sure that all but one edge in our construction are definitely separator edges.

\begin{mythm}\label{thm:npc}
It is NP-complete to decide whether a given simple drawing of an arbitrary graph is separable.
\end{mythm}

\begin{proof}
Given a 3-SAT formula $\phi$ which is an instance of linked planar 3-SAT with negated edges on one side, we construct a simple drawing $\mathcal{D}$ containing a special edge $e$ such that $e$ is a separator edge if and only if $\phi$ is satisfiable.
Therefore, it is NP-hard to decide whether $e$ is a separator edge. Moreover, we show that all other edges in $\mathcal{D}$ are definitely separator edges. Consequently, it is NP-hard to decide whether $\mathcal{D}$ is separable.
In \Cref{fig:gadget} we illustrate the individual gadgets of the following construction and in \Cref{fig:3sat-formula} we show an example of the whole drawing $\mathcal{D}$ corresponding to the embedding $G_\phi \cup \mathcal{C}$ based on a small 3-SAT formula.

\begin{figure}[htb]
\centering
\subcaptionbox{\label{fig:start-gadget}}[.16\textwidth]{\includegraphics[page=4]{Figures/np-hardness.pdf}}
\subcaptionbox{\label{fig:variable-gadget}}[.15\textwidth]{\includegraphics[page=1]{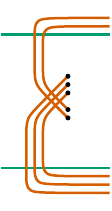}}
\subcaptionbox{\label{fig:center-gadget}}[.09\textwidth]{\includegraphics[page=5]{Figures/np-hardness.pdf}}
\subcaptionbox{\label{fig:clause-gadget-1}}[.22\textwidth]{\includegraphics[page=2]{Figures/np-hardness.pdf}}
\subcaptionbox{\label{fig:clause-gadget-2}}[.22\textwidth]{\includegraphics[page=3]{Figures/np-hardness.pdf}}
\subcaptionbox{\label{fig:end-gadget}}[.12\textwidth]{\includegraphics[page=6]{Figures/np-hardness.pdf}}
\caption{The variable gadget \subref{fig:variable-gadget} and the two clause gadgets \subref{fig:clause-gadget-1} and \subref{fig:clause-gadget-2}. Boundary edges are drawn green, literal edges darkorange, auxiliary edges purple, and local edges lightblue.}
\label{fig:gadget}
\end{figure}

Given an embedding of the union of the incidence graph $G_{\phi}$ and the Hamiltonian cycle~$\mathcal{C}$, we let $e$ be the center part of one of the two edges of $\mathcal{C}$ between the clause vertices and the variable vertices.
We then add four \emph{boundary} edges, close to $\mathcal{C}$ and on both sides next to the variable and the clause part each, crossing $e$ and crossing each other in the middle; see \Cref{fig:start-gadget,fig:center-gadget,fig:end-gadget}.
Thereby we restrict the potential witness $\gamma_e$ of $e$ to be drawn between these boundary edges, that is, within a strip close to~$\mathcal{C}$.

For each edge of $G_{\phi}$ we also add an edge to~$\mathcal{D}$, crossing the two boundary edges on its respective side.
We call these edges \emph{literal} edges.
Instead of the variable vertices of $G_{\phi}$ we let the incident literal edges in $\mathcal{D}$ cross in a grid such that edges for positive literals are drawn in one direction and those for negative literals in the other direction; see \Cref{fig:variable-gadget}.
This is possible because $\mathcal{C}$ splits those edges into inside and outside, respectively.
In that way we force $\gamma_e$ to cross either all positive or all negative literal edges of the corresponding variable in the vicinity of the variable gadget.
(Note that, to pass the variable gadget, $\gamma_e$ could also cross more literal edges, but never a proper subset of both the negative and the positive literal edges.)
Consequently, crossing the positive side  encodes the variable being set to FALSE and vice versa.

For the clause variables of $G_{\phi}$ we construct special clause gadgets depending on how many positive/negative literals are in the clause.
We can assume, for simplicity and without loss of generality, that all clauses contain exactly three literals (duplicating one literal if necessary).
Hence we have two cases, either all literals are of the same type (negated or not) or two are of one type and one of the other.
\Cref{fig:clause-gadget-1,fig:clause-gadget-2} show the two respective constructions of clause gadgets.
In addition to the literal edges, we need some \emph{auxiliary} edges that cross $e$ and a boundary edge, and some \emph{local} edges in the gadgets.
Since auxiliary edges cross~$e$, they cannot be crossed by $\gamma_e$ within the clause gadgets.
Further, literal edges can only be crossed if they were not yet crossed in the variable gadget, that is, if they have the value TRUE.
Finally, $\gamma_e$ can pass through a clause gadget without crossing any local edge twice if and only if it can cross at least one literal edge.
Hence, if there is a variable assignment satisfying~$\phi$, then we can pass the variable gadgets on the respective (other) sides and still pass all the clause gadgets to get a witness~$\gamma_e$.
Similarly, if we can find a witness~$\gamma_e$, then the way $\gamma_e$ crosses the variable gadgets gives us a valid variable assignment satisfying~$\phi$ (if~$\gamma_e$ crosses all positive literal edges within a variable gadget, then setting that variable to FALSE is valid and vice versa).

\begin{figure}[htb]
\centering
\subcaptionbox{\label{fig:3sat-incidence-graph}}{\includegraphics[page=8]{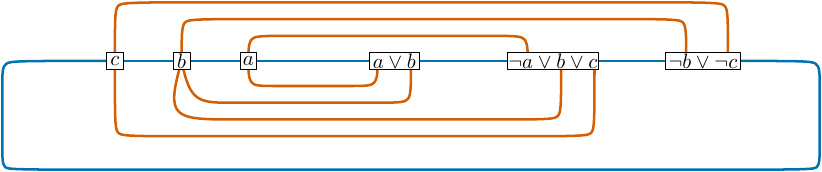}}
\subcaptionbox{\label{fig:3sat-simple-drawing}}{\includegraphics[page=7]{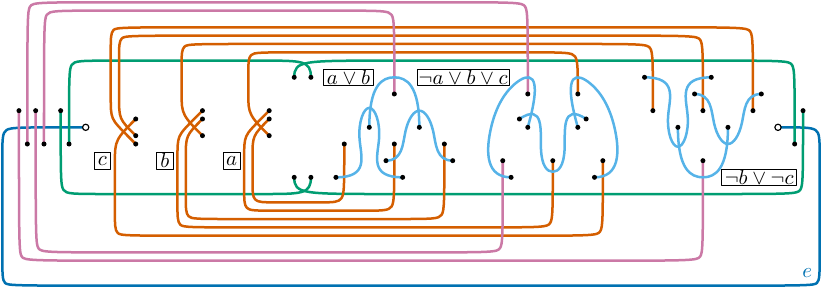}}
\caption{\subref{fig:3sat-incidence-graph}~An embedding of $G_\phi \cup \mathcal{C}$ for the 3-SAT formula $(a \vee b) \wedge (\neg a \vee b \vee c) \wedge (\neg b \vee \neg c)$. Edges of $\mathcal{C}$ are darkblue and edge of $G_\phi$ are darkorange. \subref{fig:3sat-simple-drawing}~The corresponding simple drawing $\mathcal{D}$ as an instance to decide whether the edge~$e$ (darkblue) is a separator edge. The first and third clause use the same gadget, just upside down, and the literals $a$ and $\neg c$, respectively, are duplicated to have exactly 3 literals in all clauses.}
\label{fig:3sat-formula}
\end{figure}

So far we have shown that $e$ is a separator edge if and only if $\phi$ is satisfiable.
It remains to show that all other edges in the construction are separator edges in any case.
The local edges can be closed locally within their gadget (crossing only edges that are part of the gadget).
The boundary edges can be closed next to the boundary edge on the other side of the strip because they did not cross any literal or auxiliary edges there yet.
Further, we let the auxiliary edges in the construction cross $e$ in reverse order (nested) to how they enter the strip within the boundary edges.
Therefore they pairwise do not cross and can be closed outside of the other side of the strip (potentially crossing one local and the other auxiliary edge in their clause gadget but then only crossing some literal edges directly after leaving the strip).
For the literal edges we have to be a bit more careful:
We go back next to the boundary on the other side of the strip (potentially crossing one local edge to get there).
Since $G_{\phi}$ is planar we can cross all other literal edges, except for those corresponding to the same variable.
However, in each clause gadget we can cross (potentially in addition to some local edges) either the auxiliary edge or at least one of the three literal edges on the respective side (we can assume, without loss of generality, that no clause consists of three times the same literal because then we could simplify~$\phi$).
Finally, between the clause and variable gadgets (\Cref{fig:center-gadget}) we change sides and then cross all literal edges on the same side (which we made sure to not cross already in any of the clause gadgets).

This finishes the proof for NP-hardness. For NP-completeness observe that a witness set for~$\mathcal{D}$ can be encoded and checked in polynomial space and time.
\end{proof}

\begin{figure}[htb]
\centering
\subcaptionbox{\label{fig:witness-multi-crossings}}[.53\textwidth]{\includegraphics[page=1]{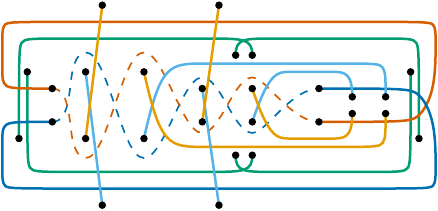}}
\subcaptionbox{\label{fig:non-separable-matching}}[.20\textwidth]{\includegraphics[page=2]{Figures/recognize-examples.pdf}}
\subcaptionbox{\label{fig:cross-triangle-3-times}}[.24\textwidth]{\includegraphics[page=3]{Figures/recognize-examples.pdf}}
\caption{\subref{fig:witness-multi-crossings}~A separable drawing where the witnesses (dashed) of the two dark edges (orange and blue) have to cross multiple times because they need to avoid (in addition to the green boundary edges) all light edges of the respective color. \subref{fig:non-separable-matching} and \subref{fig:cross-triangle-3-times}~The orange edge is not a separator edge. The latter case of an edge crossing all 3 sides of a triangle is particularly interesting because it easily explains why the twisted drawing of $K_5$ is not separable.}
\label{fig:recognize-examples}
\end{figure}

One of the differences between pseudospherical and separable drawings that we observed early on was that in the latter the witnesses of two incident edges can touch (instead of cross) in their common vertex.
This allows for significantly more freedom and one might ask whether it is all the additional freedom, that is, whether more than two crossings between witnesses can always be avoided.
Inspired by our NP-hardness construction, \Cref{fig:witness-multi-crossings} shows an example of a separable drawing on $n$ vertices where the witnesses of two of its edges have to cross linearly often.

To round off this section, \Cref{fig:non-separable-matching,fig:cross-triangle-3-times} show two small examples of non-separable simple drawings that can be easily recognized:
The end-vertices of an edge $e$ share no common cell in the subdrawing consisting of $e$ together with all edges that cross $e$ or are incident to it.

\section{Conclusion and Open Problems}\label{sec:conclusion}

In \Cref{sec:extendability} we showed that both separable and crossing-minimizing drawings of arbitrary graphs can be extended to simple drawings of complete graphs.
Recall that our research was actually triggered by the question whether all crossing-minimizing drawings of $K_n$ are pseudospherical~\cite{amrs-2022-cdcgtmg,ars-2021-edcgap}.
Therefore we would like to extend this question (in modified form) to arbitrary graphs.

\begin{myq}
Is every crossing-minimizing drawing of an arbitrary graph separable?
\end{myq}

Regarding Hamiltonicity and especially \Cref{obs:deg-2-separable}, we are curious how far our proofs of \Cref{thm:paths-separable,thm:cycle-separable} can be extended.
In particular, Harborth and Mengersen~\cite{hm-1992-dcgmnc} asked whether every crossing-maximizing drawing of $K_n$, that is, simple drawing with $\binom{n}{4}$ crossings contains at least two completely uncrossed edges.
If this can be answered to the positive, then crossing-maximizing drawings of $K_n$ definitely have some separator edges.
Moreover, the twisted drawing on $n$ vertices $v_1, \ldots, v_n$ defined such that two edges $\{ v_i, v_j \}$, $\{ v_k, v_\ell \}$ cross if and only if $i < k < \ell < j$~\cite{hm-1992-dcgmnc,pst-2003-ucctg} is degree-2-separable.
Indeed, the separator edges are exactly all edges $\{ i, i+1 \}$ for $1 \leq i < n$, the edge $\{ 1, 3 \}$, and the edge $\{ n-2, n \}$.
Therefore we ask the following, where a positive answer would also settle the Hamiltonicity question for g-twisted drawings~\cite{agtvw-2024-twfpssdcg}.

\begin{myq}
Is every crossing-maximizing drawing of $K_n$ degree-2-separable?
\end{myq}

We have given an example of a separable drawing of $K_n$ where $1/4$ of all spanning stars are not compatible with any crossing-free Hamiltonian cycle.
This is a clear difference to g-convex drawings where all spanning stars are compatible.
For simple drawings there exist examples where no spanning star is compatible with any crossing-free Hamiltonian cycle \cite[Figure~21]{bfmos-2025-phccd}.
We wonder whether this is already possible for separable drawings.

\begin{myq}
Does there exist a separable drawing of $K_n$ where no spanning star is compatible with any crossing-free Hamiltonian cycle?
\end{myq}

Another generalization of crossing-free Hamiltonian cycles considered in~\cite{bfmos-2025-phccd} are empty $k$-cycles.
As they showed, a single spanning star that is compatible with a crossing-free Hamiltonian cycle is sufficient to guarantee empty $k$-cycles for every~$k$.
Also $x$-monotone drawings clearly contain empty $k$-cycles for every $k$, for example, by considering the $k$~left-most vertices.
For separable drawings this problem remains unresolved.
However, note that if we could guarantee for every $k$ a separator edge with exactly $k$ vertices on one side, then we would also get the property.

\begin{myq}
Does every separable drawing of $K_n$ contain an empty $k$-cycle for every $3 \leq k \leq n$?
\end{myq}

Recently Bergold, Orthaber, Scheucher, and Schr{\"o}der~\cite{boss-2025-hcsd} showed that g-convex drawings of $K_n$ always contain $6$-holes, given that $n$ is large enough.
Since separable drawings of $K_n$ do not contain the twisted drawing of $K_5$, this result could potentially be further generalized.

\begin{myq}
Do all separable drawings of $K_n$, for large enough $n$, contain $6$-holes?
\end{myq}

In \Cref{sec:flips} we have introduced flips in rotation systems.
Further research in this direction could be to study properties of the flip graph.
In the flip graph every realizable rotation system is a vertex and two vertices are connected by an edge if the respective rotation systems can be transformed into each other by a single flip.
By computations, we know that the flip graph is connected for $n \leq 9$.
On the other hand, it is unclear whether every simple drawing of $K_n$ even has at least one flippable edge, that is, whether there could be isolated vertices in the flip graph.
In particular, one of the two non-realizable rotation systems of $K_5$ does not have a single flippable edge.

\begin{myq}
Is the flip graph of realizable rotation systems of size $n$ connected?
\end{myq}

Further, we showed NP-hardness for recognizing separable drawings of arbitrary graphs.
To the best of our knowledge, the corresponding question for pseudospherical drawings~\cite{ars-2021-edcgap} is still open.
The issue in generalizing our NP-hardness construction from separable to pseudospherical drawings is mostly the witness of the edge $e$, which can in general cross other witnesses an arbitrary number of times.
However, since being pseudospherical is a property of the whole drawing and not just one edge at a time, a completely different approach might work there.

\begin{myq}
What is the computational complexity of deciding whether a simple drawing of an arbitrary graph is pseudospherical?
\end{myq}

\nocite{*}
\bibliographystyle{plainurl}
\bibliography{literature.bib}

\end{document}